\documentclass[11pt]{article}

\usepackage{subeqn}
\usepackage{amssymb, amsbsy}
\usepackage{epic, eepic}
\usepackage{texdraw}
\usepackage{latexsym}

\usepackage{accents}

\newcommand{\field}[1]{\mathbb{#1}}
\title{Reduction groups and related integrable difference systems of NLS type}
\author{S. Konstantinou-Rizos$^{1,2}$,  A.V. Mikhailov$^{1}$ and P. Xenitidis$^{1}$ \\ $^{1}$ School of Mathematics, University of Leeds, LS2 9JT Leeds, UK \\ $^{2}$ Faculty of Maths \& Computer Technology, Chechen State University, 364907 Grozny, Russia}
\date{\today}

\begin{document}

\maketitle
\begin{abstract}

We extend the reduction group method to the Lax-Darboux schemes associated with 
nonlinear Schr\"odinger type equations. We consider all possible finite 
reduction groups and construct corresponding Lax operators, 
Darboux transformations, hierarchies of integrable 
differential-difference equations, integrable partial difference systems and 
associated scalar partial difference equations.
\end{abstract}

\section{Introduction}

In the theory of integrable systems the connections between partial differential
equations, differential-difference and partial difference systems are well
known. A clear and seminal account of these connections can be found in 
\cite{A-thesis}, \cite{Adler-Yamilov}.  They can be formulated in the 
frame of a Lax-Darboux scheme, where 
\begin{itemize}
 \item the Lax structure (Lax representation, also known as zero curvature representation) is associated with partial differential equations (PDEs) and their symmetries \cite{zmnp, abseg};
 
\item Darboux transformations, which are automorphisms of the Lax structure, 
lead to B\"acklund transformations which can be regarded as
integrable differential-difference equations (D$\Delta$Es) \cite{Levi-Ben80, Levi-1981,  Adler-Yamilov};

\item Bianchi permutability of the Darboux transformations yields integrable 
partial difference  equations (P$\Delta$Es) whose symmetries are the former D$\Delta$Es 
\cite{QNCL,  A}.
\end{itemize}

In this paper we extend  the reduction group method  \cite{Mikhailov2} to  
Lax-Darboux schemes for nonlinear Schr\"odinger type equations. More precisely,  we study  Lax operators of the form
\begin{equation} \label{L-operator}
{\cal{L}}\, =\, D_{x}\, +\, U(p,q;\lambda), 
\end{equation}
where  the $2 \times 2$ matrix $U$ belongs to the Lie algebra
${\mathfrak{sl}}_2({\mathbb{C}(\lambda)})$. Matrix $U(p, q;\lambda)$  
depends implicitly on $x$ through two
potentials $p$,  $q$,  and is a rational function in the spectral 
parameter
$\lambda$.  Imposing the invariance of operator $\cal{L}$ under the action of a 
reduction group, which is a finite subgroup of the group of 
automorphisms of ${\mathfrak{sl}}_2({\mathbb{C}(\lambda)})$, 
we construct systematically the Lax operators corresponding to deep reductions. 
In this case there 
is a complete classification of finite reduction groups  \cite{Bury,  Lomb-Mikh, 
 Lomb-Sand} and corresponding 
reduced Lax operators \cite{Bury}. Namely, in the 
${\mathfrak{sl}}_2({\mathbb{C}(\lambda)})$ case there are only five distinct 
cases: 
\begin{enumerate}
 \item[(i)]the trivial reduction group (no reductions); 
 \item[(ii)]  $\mathbb{Z}_2$ group with a degenerate orbit; 
\item[(iii)] $\mathbb{Z}_2$ group with a generic orbit; 
\item[(iv)]  $\mathbb{Z}_2\times \mathbb{Z}_2$ group with a degenerate orbit; 
\item[(v)] $\mathbb{Z}_2\times \mathbb{Z}_2$ group with a generic orbit. 
\end{enumerate}
 In the cases (i)-(iv) we construct an invariant Lax operator, a
corresponding PDE, invariant  Darboux transformations, corresponding integrable D$\Delta$Es and P$\Delta$Es.
The simplest case (i) has been studied in detail in \cite{Adler-Yamilov}.
We present it here for completeness, in order to 
illustrate all elements of the corresponding Lax-Darboux scheme, such as 
dressing chains (also known as 
B\"acklund transformations) and their first integrals; to give a detailed derivation 
of associated integrable P$\Delta$Es, and to discuss possible initial-value 
problems for these P$\Delta$Es. The case (v) can be studied by the methods 
presented in the paper but leads to cumbersome  expressions, and we have 
decided to omit it in order to keep our results presentable.

Darboux transformations are automorphisms of the Lax structure and discrete 
symmetries of the corresponding PDEs. With each Darboux transformation we 
associate an 
infinite lattice and a map. If there are two Darboux transformations, then the 
condition 
of their commutativity (the Bianchi permutability) yields 
an integrable system of P$\Delta$Es. 

Although the theory of Darboux transformations is rather well developed and has 
a long history, there are a few important problems which require further 
research. One of the problems is to give a complete description of all possible 
Darboux transformations for a given Lax operator. In the case of the 
Schr\"odinger operator the solution is known: there is one Darboux transformation (depending on a parameter) and any other Darboux 
transformation can be represented as a composition of such 
transformations and their inverses for a certain choice of the 
parameters \cite{A-thesis}. However the description of all 
possible Darboux transformations associated with a given Lax 
operator is still an open problem.

The paper is organised as follows. In the following section,  we introduce our
notation and give the general scheme of these considerations. In the next four
sections we consider the Lax operators related to the nonlinear Schr{\"o}dinger
equation (Section \ref{sec-NLS}),  and operators derived from the reduction
group method,  \cite{Lomb-Sand,  Bury},  namely ${\mathbb{Z}}_2$ reduction
(Sections \ref{sec-Z2-deg} and \ref{sec-Z2-gen}) and dihedral group reduction
(Section \ref{sec-Dih}).

\section{Lax-Darboux scheme} \label{LaxOp-Darb}

In this section,  we explain our terminology by describing the Lax-Darboux scheme. We present the class of Lax operators under consideration and discuss our general assumptions for the construction of Darboux matrices. Moreover, we introduce the notation we use throughout the paper.

With the single  term  {\emph{Lax-Darboux scheme}} we describe several 
structures which are related to each other and all of them are related to 
integrability. To be more precise, the Lax-Darboux scheme incorporates Lax 
operators, corresponding Darboux matrices and Darboux transformations, as well 
as the Bianchi permutability of the latter transformations.
\begin{itemize}
\item Lax operators are linear operators of the form $ {\cal{L}} = D_x + 
U$, where the $N\times N$ matrix $U$ is an element of a specific Lie algebra.  
As it was described in the previous section, in this paper we consider only the 
case where $U(p,q;\lambda)$ is a $2 \times 2$ matrix belonging to the Lie 
algebra ${\mathfrak{sl}}_2({\mathbb{C}(\lambda)})$, and its dependence on the 
continuous variable $x$ is implicit through the potentials $p$ and $q$.

\item Darboux transformations ${\cal{S}}$ are automorphisms of  the Lax operator ${\cal{L}}$. They map ${\cal{L}}$ to $\widetilde{\cal{L}}$ by updating potentials $p$ and $q$. In other words, 
$${\cal{S}} : {\cal{L}} \mapsto \widetilde{\cal{L}},\quad {\mbox{where}}\quad 
{\cal{L}} = D_x + U(p,q;\lambda), \quad \widetilde{\cal{L}} =  D_x + 
U(\widetilde{p},\widetilde{q};\lambda),$$
with $\widetilde{p}$, $\widetilde{q}$ denoting the updated potentials. 
 
Darboux transformations consist of Darboux matrices $M$ along with corresponding dressing chains or B{\"a}cklund transformations. 

\item A Darboux matrix $M$ maps a fundamental solution of the equation 
${\cal{L}}(\Psi) =0$ to a fundamental solution $\widetilde{\Psi}$ of 
$\widetilde{\cal{L}}(\widetilde{\Psi}) =0$ according to $\widetilde{\Psi} = M 
\Psi$. In general, matrix $M$ is invertible and depends on $p$, $q$, their 
updates $\widetilde{p}$, $\widetilde{q}$, the spectral parameter $\lambda$, and 
some auxiliary functions.

\item Dressing chains are sets of differential equations relating the 
potentials and the auxiliary functions involved in ${\cal{L}}$ and 
$\widetilde{\cal{L}}$. They can be regarded as integrable systems of 
D$\Delta$Es.  This follows from the interpretation of the corresponding Darboux 
transformation as defining a shift on the lattice according to the sequence
$$\cdots \ \stackrel{{\cal{S}}}{\longrightarrow}\
(\undertilde{p},\undertilde{q}) \ \stackrel{{\cal{S}}}{\longrightarrow} 
\ (p,q) \  \stackrel{{\cal{S}}}{\longrightarrow} \ 
(\widetilde{p},\widetilde{q}) \ 
\stackrel{{\cal{S}}}{\longrightarrow} \  \cdots.$$

\item If the Lax operator admits two commuting Darboux transformations 
${\cal{S}}$ and ${\cal{T}}$, then they define a two-dimensional lattice for which 
we adopt the multi-index notation $(p_{ij},q_{ij})={\cal{S}}^i 
{\cal{T}}^j(p,q)$, where $i, j \in {\mathbb{Z}}$.   
This interpretation allows us to derive systems of integrable 
P$\Delta$Es by considering the Bianchi permutability of  the corresponding 
transformations.

\end{itemize}

In order to implement the above scheme, firstly we construct Darboux 
transformation ${\cal{S}}$. From the definition of Darboux matrix $M$ 
follows that
\begin{equation}\label{trans}
M {\cal{L}} M^{-1}\, =\, \widetilde{{\cal{L}}}\,,
\end{equation}
or, denoting the updated potentials with $p_{10}$, $q_{10}$ and matrix 
$U(p_{10},q_{10};\lambda)$ with $U_{10}$, we can rewrite equation 
(\ref{trans}) explicitly as
\begin{equation} \label{compatibility1}
D_x M\, +\,  U_{10}M - M U\, =\, 0\,.
\end{equation}
For a given Lax operator $\cal{L}$, the above equation can be used to determine 
$M$, as well as the corresponding dressing chain. Moreover, since matrices $U$ 
and $U_{10}$ are traceless, it follows from Abel's theorem that the determinant 
of $M$ is a first integral of the dressing chain. For the Lax operators we 
consider here, it is natural to assume that matrix $M$ depends rationally on the 
spectral parameter $\lambda$, and inherits the reduction group symmetries of the 
corresponding operator $\cal{L}$.

The interpretation of the Darboux transformation $\cal{S}$ as defining a lattice direction 
allows us to think the updated potentials in $\widetilde{\cal{L}}$ as shifts of 
the original ones $p$, $q$ in that particular lattice direction.  In this 
semi-discrete setting,  the corresponding dressing chain can be seen as an 
integrable differential-difference equation \cite{Levi-Ben80, Levi-1981} 
deriving from the compatibility condition of the {\emph{Lax-Darboux pair}} 
(also referred to as semi-discrete Lax pair)
$$ D_x \Psi = - U(p,q;\lambda)\Psi, \quad \Psi_{10} = M(p,q,p_{10},q_{10};\lambda) \Psi\,.$$

In this discrete interpretation, the Bianchi permutability of two different Darboux transformations yields an integrable system of P$\Delta$Es in two discrete variables. Employing the standard notation for difference equations, we denote the two discrete variables with $n$ and $m$, and interpret $\cal{S}$ and $\cal{T}$ as the corresponding shift operators defined by 
$${\cal{S}}^i{\cal{T}}^j \left(h(n, m)\right) = h(n+i, m+j) \equiv h_{ij}\,.$$
In particular, when $i=j=0$,  we will omit the index ``00'', i.e. $h = h(n,m)$.

\begin{figure}[ht]
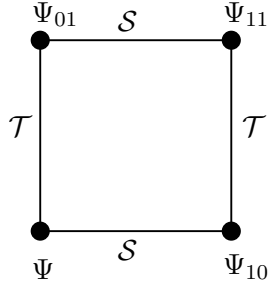

\centertexdraw{ \setunitscale 0.5
\linewd 0.02 \arrowheadtype t:F 
\htext(0 0.5) {\phantom{T}}
\move (-1 -2) \lvec (1 -2) 
\move(-1 -2) \lvec (-1 0) \move(1 -2) \lvec (1 0) \move(-1 0) \lvec(1 0)
\move (1 -2) \fcir f:0.0 r:0.1 \move (-1 -2) \fcir f:0.0 r:0.1
 \move (-1 0) \fcir f:0.0 r:0.1 \move (1 0) \fcir f:0.0 r:0.1  
\htext (-1.1 -2.5) {$\Psi$} \htext (.9 -2.5) {$\Psi_{10}$} \htext (-.2 -2.3)
{$\cal{S}$}
\htext (-1.1 .15) {$\Psi_{01}$} \htext (.9 .15) {$\Psi_{11}$} \htext (-.2 .1)
{$\cal{S}$}
\htext (-1.35 -1) {$\cal{T}$} \htext (1.1 -1) {$\cal{T}$}}
\caption{{\em{Bianchi commuting diagram}}} \label{fig:bianchi}
\end{figure}

Now the shift operators $\cal{S}$ and $\cal{T}$ act on a fundamental solution $\Psi$ as
\begin{equation} \label{dis-LP}
\begin{array}{l}
{\cal{S}} : \Psi \mapsto \Psi_{10}\, =\, M(p,q,p_{10},q_{10};f;\lambda)\Psi \equiv
M\, \Psi\, , \\
{\cal{T}} : \Psi \mapsto \Psi_{01}\, =\,  K(p,q,p_{01},q_{01};g;\lambda) \Psi \equiv K\, \Psi,\end{array}
\end{equation}
where $M$ and $K$ are the corresponding Darboux matrices with $f$ and $g$ denoting any auxiliary (vector) functions. 
The Bianchi permutability of (\ref{dis-LP}) according to Figure \ref{fig:bianchi} allows us to compute $\Psi_{11}$ in two different ways. This 
yields the consistency condition
\begin{equation} 
{\cal{T}}(M)\, K\, -\, {\cal{S}}(K)\, M\, =\, 0, \label{comDT-LP}
\end{equation}
which is nothing else but the compatibility condition of the {\emph{Darboux 
pair}} (also referred to as fully discrete Lax pair)
$${\cal{S}}(\Psi) = M(p,q,p_{10},q_{10};f;\lambda) \Psi\,,\quad {\cal{T}}(\Psi) = K(p,q,p_{01},q_{01};g;\lambda) \Psi\,. $$

The resulting condition (\ref{comDT-LP}) yields a set of polynomial equations 
for $p$, $q$, $f$, $g$ and their shifts. This set may have two branches of 
solutions. One of them leads to a trivial system, cf. (\ref{triv-sol}) below,  
whereas the other branch yields a non-trivial integrable system of partial 
difference equations. Symmetries and first integrals for the non-trivial 
system follow from the dressing chain and the first integrals of the 
corresponding Darboux transformations.

For some of these discrete systems, we employ first integrals and conservation 
laws to reduce the number of dependent variables and derive integrable scalar 
equations of Toda type. The form of these systems allows us to formulate a 
Cauchy problem on a single or a double staircase. 

In our derivations,  we find more than one Darboux transformation for each
Lax operator we consider. We would like to emphasise here that the
interpretation of any pair of Darboux matrices as a discrete Lax pair as 
described above does not always lead to a non-trivial discrete 
system. In the following sections we present only the pairs of Darboux 
matrices which lead to genuinely non-trivial discrete integrable systems.

\section{Nonlinear Schr{\"o}dinger equation} \label{sec-NLS}

In order to illustrate our approach,  we consider a well known operator
\begin{equation} \label{NLS-U}
{\cal{L}} \ = \ D_x + U(p, q;\lambda)\, =\, D_x + \lambda\,  \sigma_3\, +\,
\left(\begin{array}{cc} 0 & 2\, p \\ 2\, q & 0\end{array}\right),
~~~\sigma_3={\rm{diag}}(1, -1), 
\end{equation}
which is the spatial part of the Lax pair for the nonlinear Schr\"odinger
equation \cite{ZS}
\begin{equation} \label{NLS}
p_t=p_{xx}+4\, p^2\, q, ~~~q_t=-q_{xx}-4\, p\, q^2.
\end{equation}

It is straightforward to verify that the constant matrix
$$M = \left(\begin{array}{cc} \alpha & 0 \\ 0 & \beta \end{array} \right), \quad
\alpha \beta \ne 0, $$
is a Darboux transformation for this operator corresponding to the
scaling symmetry of (\ref{NLS}).
$$ p_{10} = \alpha \beta^{-1} p, \quad q_{10} = \beta \alpha^{-1} q.$$
The simplest $\lambda$-dependent Darboux matrix one may consider is
\begin{equation} \label{NLS-M}
M=\lambda M_{1}+M_{0} .
\end{equation}
Substituting (\ref{NLS-U}) and (\ref{NLS-M}) into the compatibility condition
(\ref{compatibility1}),  the coefficient of $\lambda^2$ implies that matrix
$M_{1}$ must be diagonal. Additionally,  from the diagonal part of the
coefficient of $\lambda$ we conclude that $M_1$ must be constant. Hence,  $M_1 =
{\rm{diag}} (c_1, c_2)$. We could choose either $c_1=1$,  $c_2=0$,  or $c_1=0$, 
$c_2=1$ or $c_1=c_2=1$. Since the  first two choices are gauge equivalent and the
third one can be given as a composition of two suitable  Darboux
matrices with $c_1 c_2 = 0$,  we choose
$$M_1 = \left( \begin{array}{cc} 1 & 0 \\ 0 & 0\end{array} \right). $$
Moreover,  the off-diagonal part of the coefficient of $\lambda$ implies that
the (2, 2) element of $M_0$ is constant and,  hence,  we have to consider two
distinct cases.

The first case corresponds to
$$M  = \lambda \left(\begin{array}{cc} 1 & 0 \\ 0 & 0 \end{array} \right)  +
\left(\begin{array}{cc} f & a \\ b & 0\end{array} \right), $$
i.e. the $(2, 2)$ entry of $M_0$ is zero. In this case,  equation
(\ref{compatibility1}) is equivalent to the system 
\begin{equation}
a  = p,  \qquad b = q_{10}, \qquad f_x=2\, (a\, q-b\, p_{10}),  \qquad a_x=2\,
f\, p,  \qquad b_x=-2\, f\, q_{10}.
\end{equation}
The first two equations determine functions $a$ and $b$,  while the last two 
implies that $pq_{10}=\gamma$, where $\gamma$ is a non-zero constant (since 
$\det M=\gamma\ne 0$). Without any loss of generality we can set $\gamma=1$ 
and thus we have 
\begin{equation} \label{NLS-deg-DM-BT}
q_{10}= \frac{1}{p}, \quad  p_{10}=p\, \left(p\, q-\frac{1}{2} f_x\right), \quad
f\, =\, \frac{p_x}{2\, p}.
\end{equation}
Finally, the Darboux matrix  is given by 
\begin{equation}\label{M-NLS2}
M(p, f) = \lambda \left(\begin{array}{cc} 1 & 0\\ 0 & 0
\end{array}\right)+\left(\begin{array}{cc} f & p\\ \frac{1}{p} & 0 \end{array}
\right), 
\end{equation}
and the dressing chain (the B\"acklund transformation  (\ref{NLS-deg-DM-BT})) 
can be rewritten in the form of the Toda lattice in a new variable $\phi=\log p$
\[
 \phi_{xx}=4 e^{\phi-\phi_{-1,0}}-4e^{\phi_{10}-\phi}.
\]
In this case the Darboux transformation $(p,q)\to (p_{10},q_{10})$ is explicit
\[
 p_{10}=p\left(pq-\frac{1}{4}\left(\frac{p_x}{p}\right)_x\right),\qquad 
q_{10}=\frac{1}{p}.
\]

Alternatively,  we can choose the $(2, 2)$ element of $M_0$ to be non zero and, 
without loss of generality set it to 1,  i.e.
$$M_0 =  \left( \begin{array}{cc} f & a \\ b & 1 \end{array} \right).$$
Now, it follows from (\ref{compatibility1})
that
\begin{subequations} \label{nls-comp-cond}
\begin{eqnarray}\nonumber
&& a\, =\, p\, , \quad b\, =\, q_{10}\, , \\
&& \partial_x f\, =\,  2 (pq-p_{10}\, q_{10})\, , \\
&& \partial_x p \, =\, 2 (p f -p_{10})\, , \quad \partial_x q_{10}\, =\, 2 (
q-q_{10}\, f)\, .
\end{eqnarray}
\end{subequations}
A first integral of the above system is provided by the determinant of  $M$, 
$\det M=\lambda+f-pq_{10}$
\begin{equation} \label{nls-const}
\partial_x \left(f-p\, q_{10} \right)\, =\, 0\, .
\end{equation}
Hence,  matrix $M$ has the following form
\begin{equation}
M(p, q_{10}, f) =\lambda \left(\begin{array}{cc}1 & 0\\0 &
0\end{array}\right)+\left(\begin{array}{cc}f & p\\q_{10} & 1 \end{array}\right)
\label{M-NLS}
\end{equation}
and (\ref{nls-comp-cond}) is the corresponding dressing chain.

\subsection{Derivation of discrete systems}

Having derived two Darboux matrices for operator (\ref{NLS-U}),  we focus on the
generic one given in (\ref{M-NLS}) and consider the following Darboux pair
$$\Psi_{10} = M(p, q_{10}, f) \Psi, \quad \Psi_{01} = M(p, q_{01}, g) \Psi, $$
which explicitly reads as follows.
\begin{equation} \label{NLS-disc-LP}
\Psi_{10} = \left(\lambda \left(\begin{array}{cc} 1 & 0\\0 & 0
\end{array}\right)+\left(\begin{array}{cc} f & p\\q_{10} &
1\end{array}\right)\right) \Psi, \quad \Psi_{01} = \left( \lambda
\left(\begin{array}{cc} 1 & 0\\ 0 & 0\end{array}\right)+\left(\begin{array}{cc}
g & p\\ q_{01} & 1\end{array}\right) \right)\Psi.
\end{equation}
The compatibility condition of (\ref{NLS-disc-LP}) results to
\begin{subequations} \label{nls-comp}
\begin{eqnarray}
&& f_{01} -f - \left( g_{10}-g\right) = 0\, , \\
&& f_{01}\, g-f\, g_{10}\, -\, p_{10}\, q_{10}\, +\, p_{01}\, q_{01}\, =\, 0\, ,
\\
&& p \left(f_{01}-g_{10} \right)\, -\, p_{10}+p_{01}\, =\, 0\, , \\
&& q_{11}\, \left(f-g\right)\, -\, q_{01}+q_{10}=0\, .
\end{eqnarray}
\end{subequations}
This system can be solved either for $(p_{01}, q_{01}, f_{01}, g)$ or for
$(p_{10}, q_{10}, f, g_{10})$. It has two branches of solutions. A trivial one 
\begin{equation} \label{triv-sol}
p_{10} = p_{01}, \quad q_{10} = q_{01},  \quad f = g, \quad g_{10} = f_{01}, 
\end{equation}
corresponds to $M(p, q_{10}, f)=M(p, q_{01}, g)$, and a non-trivial solution 
given by
\begin{subequations} \label{LPcompatEq}
\begin{eqnarray}
&& p_{01} = \frac{q_{10} p^2 + (g_{10} - f) p + p_{10}}{1+p\,  q_{11}}\, ,
\quad 
q_{01} = \frac{p_{10}\, { q_{11}}^{\, 2} + (f-g_{10})\,  q_{11} + q_{10}}{1+p\, 
q_{11}}\, ,   \\
&& f_{01} = \frac{q_{11}\,  (p_{10} + p g_{10}) + f - p q_{10}}{1+ p\,  q_{11}}
\, , \quad 
g = \frac{q_{11}\,  (p f- p_{10}) + g_{10}+p q_{10}}{1+p\, q_{11}}\, . 
\end{eqnarray}
\end{subequations}
Some properties of the above system follow immediately from the derivation of
the corresponding Darboux transformations. First of all,  it admits two first
integrals,  cf. relation (\ref{nls-const}),  namely
\begin{equation}
 \label{nlsfi}
 ({\cal{T}}-1)\left(f-pq_{10}\right)=0\quad {\mbox{and}} \quad
({\cal{S}}-1)\left(g-pq_{01}\right)=0. 
\end{equation}
We can interpret functions $f$ and $g$ as
being given on the edges of the quadrilateral where system (\ref{LPcompatEq}) is
defined,  and,  consequently,  consider system (\ref{LPcompatEq}) as a
vertex-bond system \cite{HV}. System (\ref{LPcompatEq}) admits
the conservation law
$$({\cal{T}}-1)f=({\cal{S}}-1)g$$
which is the first equation in (\ref{nls-comp}).

Moreover,  relations (\ref{nls-comp-cond}) imply that system (\ref{LPcompatEq})
admits one generalised symmetry generated by the differential-difference
equations
\begin{eqnarray}
&& \partial_x p = f p - p_{10} = g p -p_{01}, \nonumber \\
&& \partial_x q = q_{-10} -f_{-10}q = q_{0, -1} - g_{0, -1}q, 
\label{sym-nls-1}\\ 
&& \partial_x f = p\,  q - p_{10}q_{10}, \quad \partial_x g =  p q - p_{01}
q_{01}.\nonumber
\end{eqnarray}

\begin{figure}[ht]
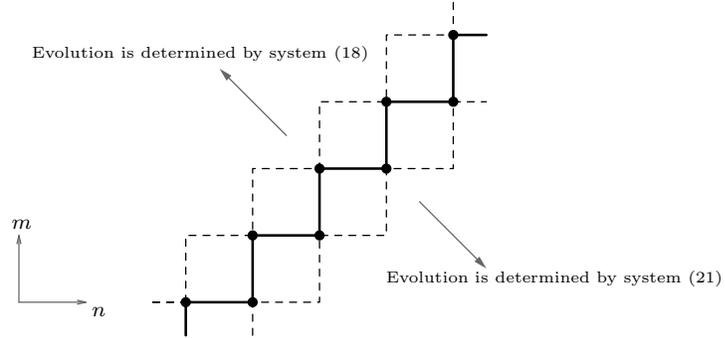

\centertexdraw{
\setunitscale 0.35
\move(-4.5 -2) \linewd 0.02 \setgray 0.4 \arrowheadtype t:V \arrowheadsize l:.12
w:.06 \avec(-4.5 -1) 
\move(-4.5 -2) \arrowheadtype t:V  \avec(-3.5 -2)
\arrowheadsize l:.20 w:.10
\move(-.5 .5) \linewd 0.02 \setgray 0.4 \arrowheadtype t:F \avec(-1.5 1.5) 
\move(1.5 -.5) \linewd 0.02 \setgray 0.4 \arrowheadtype t:F \avec(2.5 -1.5) 
\setgray 0.0
\linewd 0.04 \move (-2 -2.5) \lvec (-2 -2) \lvec (-1 -2) \lvec (-1 -1) \lvec (0
-1) \lvec (0 0) \lvec (1 0) \lvec(1 1) \lvec (2 1) \lvec(2 2) \lvec(2.5 2)
\linewd 0.015 \lpatt (.1 .1 ) \move (-2 -2) \lvec (-2 -1) \lvec(-1 -1) \lvec (-1
0) \lvec (0 0) \lvec (0 1) \lvec(1 1) \lvec (1 2) \lvec (2 2) \lvec (2 2.5)
\move(-2.5 -2) \lvec(-2 -2) \move(-2.5 -2) \lvec(-2 -2)
\move (-1 -2.5) \lvec (-1 -2) \lvec(0 -2) \lvec(0 -1) \lvec(1 -1) \lvec(1 0)
\lvec(2 0) \lvec(2 1) \lvec(2.5 1)
\move (-2 -2) \fcir f:0.0 r:0.075 \move (-1 -2) \fcir f:0.0 r:0.075
\move (-1 -1) \fcir f:0.0 r:0.075 \move (0 -1) \fcir f:0.0 r:0.075
\move (0 0) \fcir f:0.0 r:0.075 \move (1 0) \fcir f:0.0 r:0.075  
\move (1 1) \fcir f:0.0 r:0.075 \move (2 1) \fcir f:0.0 r:0.075
\move (2 2) \fcir f:0.0 r:0.075
\htext (-3.4 -2.2) {\scriptsize{$n$}}
\htext (-4.6 -.9) {\scriptsize{$m$}}
\htext (-4.3 1.6) {{\tiny{Evolution is determined by system (\ref{LPcompatEq})}}}
\htext (1 -1.75) {{\tiny{Evolution is determined by system
(\ref{LPcompatEq-2})}}}
}
\caption{{\em{Initial value problem and direction of evolution}}}
\label{fig-ivp}
\end{figure}

Our choice to solve system (\ref{nls-comp}) for $p_{01}$,  $q_{01}$,  $f_{01}$
and $g$ is motivated by the initial value problem related to system
(\ref{LPcompatEq}). Suppose that initial values for $p$ and $q$ are given at the
vertices along the solid staircase as shown in Figure \ref{fig-ivp}. Functions 
$f$ and
$g$ are given on the edges of this initial value configuration in a consistent
way with the first integrals (\ref{nls-fi}). In particular,  horizontal edges
carry the initial values of $f$ and vertical edges the corresponding ones of
$g$.
With these initial conditions,  the values of $p$ and $q$ can be uniquely
determined at every vertex of the lattice,  while $f$ and $g$ on the
corresponding edges. This is obvious from the rational expressions
(\ref{LPcompatEq}) defining the evolution above the staircase,  cf. Figure
\ref{fig-ivp}. For the evolution below the staircase,  one has to use 
\begin{subequations} \label{LPcompatEq-2}
\begin{eqnarray} 
p_{10} &=& \frac{q_{01} p^2 + (f_{01} - g) p + p_{01}}{1+p\,  q_{11}}\, , \quad 
q_{10} = \frac{p_{01}\, { q_{11}}^{\, 2} + (g-f_{01})\,  q_{11} + q_{01}}{1+p\, 
q_{11}}\, ,\\
g_{10} &=& \frac{q_{11}\,  (p_{01} + p f_{01}) + g - p q_{01}}{1+ p\,  q_{11}}
\, , \quad 
f = \frac{q_{11}\,  (p g- p_{01}) + f_{01}+p q_{01}}{1+p\, q_{11}}\, ,
\end{eqnarray}
\end{subequations}
which uniquely defines the evolution below the staircase as indicated in Figure
\ref{fig-ivp}.

We could consider more general initial value configurations of staircases of
lengths $\ell_1$ and $\ell_2$ in the $n$ and $m$ lattice direction, 
respectively. Such initial value problems are consistent with evolutions
(\ref{LPcompatEq}),  (\ref{LPcompatEq-2}) determining the values of all fields
uniquely at every vertex and edge of the lattice.

It follows from (\ref{nlsfi}) that
\begin{equation} \label{nls-fi}
f-p\, q_{10}\, =\, \alpha(n)\quad {\mbox{and}} \quad g-p\, q_{01}\, =\,
\beta(m)\, .
\end{equation}
and we can use these relations to eliminate $f$ and $g$ from  
(\ref{LPcompatEq-2}).
This results to a non-autonomous partial difference system for $p$ and $q$
only
\begin{equation} \label{nls-pq-sys}
p_{01}\, =\, p_{10}\, -\, \frac{\alpha(n)-\beta(m)}{1+ p\, q_{11}}\, p\, , \quad
q_{01}\, =\, q_{10}\, +\, \frac{\alpha(n)-\beta(m)}{1+ p\, q_{11}}\, q_{11}\, .
\end{equation}
Symmetries of this system can be derived directly from corresponding symmetries
of system (\ref{LPcompatEq}) by taking into account (\ref{nls-fi}). In
particular, it follows from (\ref{sym-nls-1}) that
\begin{eqnarray*}
&&\partial_x p = 2 (p^2\, q_{10} + \alpha(n) p - p_{10}), \quad \partial_x q = 2
(q_{-10} -p_{-10} \, q^2 - \alpha(n-1) q)
\end{eqnarray*}
is a symmetry of (\ref{nls-pq-sys}).

\subsubsection{Derivation of the discrete Toda equation}

Returning now to the construction of a discrete Lax pair,  we
 employ matrix $M(p, f)$,  given in (\ref{M-NLS2}),  and matrix 
$M(p, q_{01}, g)$,  in (\ref{M-NLS}). That is,  we consider the
following system
$$ \Psi_{10} = \left(\lambda \left(\begin{array}{cc} 1 & 0\\0 & 0
\end{array}\right)+\left(\begin{array}{cc} f & p\\ \frac{1}{p} & 0
\end{array}\right)\right) \Psi, \quad \Psi_{01} = \left( \lambda
\left(\begin{array}{cc} 1 & 0\\ 0 & 0\end{array}\right)+\left(\begin{array}{cc}
g & p\\ q_{01} & 1\end{array}\right) \right)\Psi.$$
The compatibility condition of the above system implies that
$$ p = \frac{1}{q_{10}}, \quad g = \alpha(m) + \frac{q_{01}}{q_{10}}, $$
as well as 
$$f\, =\, \frac{q_{01}}{q_{10}} - \frac{q_{10}}{q_{11}}+\alpha(m), \quad f_{01}
= \frac{q_{11}}{q_{20}} - \frac{q_{10}}{q_{11}}+ \alpha(m).$$
From the consistency of the latter equations and setting $q = \exp(-w_{-1,
-1})$,  we derive the fully discrete Toda equation \cite{Hir, Su}
\begin{equation} \label{NLS-Toda}
{\rm{e}}^{w_{01}-w} - {\rm{e}}^{w-w_{0, -1}} + {\rm{e}}^{w_{1, -1}-w} -
{\rm{e}}^{w-w_{-1, 1}} = \alpha(m+1) -\alpha(m),  
\end{equation}
along with its generalised symmetry
$$\partial_x w\, =\,  {\rm{e}}^{w-w_{0, -1}} - {\rm{e}}^{w_{1, -1}-w} -
\alpha(m).$$
Moreover,  a conserved form of Toda equation is
$$({\cal{T}}-1)\Big({\rm{e}}^{w_{0, -1}-w_{-10}}- {\rm{e}}^{w-w_{0, -1}} +
\alpha(m) \Big)\, =\, ({\cal{S}}-1) {\rm{e}}^{w_{0, -1}-w_{-10}}.$$
It is worth noting that a staircase initial value problem for the Toda equation
(\ref{NLS-Toda}) involves the points $w_{i, -i}$ and $w_{i, -i-1}$,  i.e. a
staircase which is the reflection of the one shown in Figure \ref{fig-ivp} with
respect to a vertical or horizontal line of the discrete plane.

\section{$\field{Z}_2$ reduction group: Degenerate orbit} \label{sec-Z2-deg}

Let us now consider an operator ${\cal{L}}(\lambda)$ which is invariant under the
transformation
\begin{equation} \label{sym_cond}
s_1(\lambda): {\cal{L}}(\lambda) \rightarrow \sigma_{3}{\cal{L}}(-\lambda)\sigma_{3}.
\end{equation}
The above involution generates the reduction group \cite{Mikhailov2} which is isomorphic to the
$\field{Z}_2$ group. The invariant operator corresponding to this orbit can be
taken in the form
\begin{equation} \label{SL2-Lax-Op}
{\cal{L}}=D_x+\lambda^{2}\,  \sigma_3\, +\, \lambda \, \left(\begin{array}{cc} 0 & 2\, p
\\ 2\, q & 0\end{array}\right), 
\end{equation}
and it is the spatial part of the Lax pair for the derivative nonlinear
Schr\"odinger equation \cite{KN}
\begin{equation} \label{dNLS}
p_t=p_{xx}+4\, (p^2\, q)_x, ~~~q_t=-q_{xx}+4\, (p\, q^2)_x.
\end{equation}

It can be easily verified that the constant matrix 
\begin{equation} \label{Z2-point-sym}
M = \left(\begin{array}{cc} \alpha & 0 \\ 0 & \beta \end{array} \right), \quad
\alpha \beta \ne 0, 
\end{equation}
is a Darboux matrix for operator (\ref{SL2-Lax-Op}) corresponding
to the scaling symmetry of system (\ref{dNLS}).
$$ p_{10} = \alpha \beta^{-1} p, \quad q_{10} = \beta \alpha^{-1} q.$$

Considering Darboux matrix $M$ with the same symmetry, i.e. $M(\lambda)=\sigma_{3}M(-\lambda) \sigma_{3}$, we find after some analysis that the simplest $\lambda$-dependent Darboux matrix can be written in the form 
$$ M=\lambda^{2}M_2+\lambda M_1+M_0,$$
where matrices $M_2$ and $M_0$ are diagonal and matrix $M_1$ is off-diagonal.
Additionally,  from the compatibility condition (\ref{compatibility1}) follows that
$M_0$ is a constant matrix. Moreover, following an argument similar to the one we used in the previous section, we consider only the case ${\rm{rank}}(M_2) = 1$. Hence,  summarizing the above analysis,  we choose
$$M_2 = \left(\begin{array}{cc} f & 0 \\ 0 & 0 \end{array}\right), \quad M_1 =
\left(\begin{array}{cc} 0 & a \\ b & 0\end{array} \right) \quad
{\mbox{and}}\quad M_0 = \left(\begin{array}{cc} c_1 & 0 \\ 0 & c_2  \end{array}
\right),  \quad c_1, \, c_2\, \, \in\,  {\mathbb{C}}.$$ 

With these choices,  equation (\ref{compatibility1}) firstly determines
functions $a$,  $b$ in terms of $f$,  $p$ and $q_{10}$. In particular we find
that
$$a \, = \, f\, p, \quad b \, =\, f\, q_{10}.$$
In terms of these relations,  Darboux matrix becomes
\begin{equation} \label{DT-sl2-gen}
M(p, q_{10}, f;c_1, c_2) = \lambda^{2}\left(\begin{array}{cc} f & 0\\ 0 &
0\end{array}\right)+\lambda\left(\begin{array}{cc} 0 & f \, p\\ f\, q_{10} &
0\end{array}\right)+\left(\begin{array}{cc} c_1 & 0\\ 0 & c_2
\end{array}\right)\, , 
\end{equation}
and we derive the B{\"a}cklund transformation
\begin{eqnarray} \label{sl2-D-sym-gen}
\partial_x p &=& 2\,  p\,  \left(p_{10}\, q_{10}-p\, q \right) - 2\, \frac{c_2
p_{10} - c_1 p}{f}\, ,  \quad
\partial_x q_{10} ~=~ 2 q_{10}\, \left(p_{10}\, q_{10}-p\, q \right) - 2\,
\frac{c_1 q_{10}- c_2 q}{f},  \nonumber \\
\partial_x f &=& 2 f \left(p\, q-p_{10}\, q_{10} \right).
\end{eqnarray}
A first integral of the above system,  which also guarantees that the
determinant of matrix (\ref{DT-sl2-gen}) is independent of $x$,  is given by
\begin{equation} \label{sl2-D-con-det-gen}
\partial_x \left(f^{2}p\, q_{10}\, -\, c_2 f\right)\, =\, 0.
\end{equation}

It is apparent that if constants $c_1$,  $c_2$ are not zero,  we can always set
them to 1 by composing Darboux matrix (\ref{DT-sl2-gen}) with an appropriate Darboux matrix
(\ref{Z2-point-sym}). Hence,  we can impose without loss of generality that
these constants are either 0 or 1. There are two particular sets of values for
these constants at which differential-difference equations (\ref{sl2-D-sym-gen})
can be brought to polynomial form.
\begin{enumerate}
\item First we consider the case when $c_1=c_2=0$. It follows from equations
(\ref{sl2-D-sym-gen}) that $f = 1/p$ and $q_{10}=p$,  in view of which matrix
$M$ degenerates to
\begin{equation}\label{Z2-case1-DM-M}
 M(p) = \lambda^2 \left(\begin{array}{cc} 1/p & 0 \\ 0 & 0\end{array}
\right)\, +\, \lambda \left( \begin{array}{cc} 0 & 1 \\ 1 & 0
\end{array}\right).
\end{equation}
The corresponding B{\"a}cklund transformation becomes
 \begin{equation} \label{Z2-case1-BT}
q_{10} \, = \, p,  \quad \partial_x p\, =\, 2\, p^2\, \left(p_{10}-q \right)
\end{equation}
and the first integral (\ref{sl2-D-con-det-gen}) holds identically. The
resulting differential-difference equations (\ref{Z2-case1-BT}) are the modified
Volterra chain.

\item When $c_1=1$ and $c_2=0$,  the Darboux matrix becomes
\begin{equation} \label{Z2-case2-DM-M}
M(p, q_{10}, f) = \lambda^2 \left(\begin{array}{cc} f & 0 \\ 0 & 0
\end{array} \right) + \lambda \left(\begin{array}{cc} 0 & f p \\ f q_{10} & 0
\end{array} \right) + \left(\begin{array}{cc} 1 & 0 \\ 0 & 0\end{array}
\right)\, , 
\end{equation}
the B{\"a}cklund transformation simplifies to
\begin{equation} \label{Z2-case2-BT}
\partial_x p = 2 p \left( p_{10} q_{10}-p q\right) + \frac{2 p}{f},  \quad
\partial_x q_{10} = 2 q_{10} \left(p_{10} q_{10}-p q\right) -\frac{2 q_{10}}{f},
 \quad \partial_x f = 2 f (pq-p_{10} q_{10}), 
\end{equation}
and the first integral (\ref{sl2-D-con-det-gen}) becomes
\begin{equation} \label{Z2-case2-fi}
\partial_x\left(f^2 \, p\, q_{10}\right) = 0.
\end{equation}
In the context of differential-difference equations,  if we make the point
transformation
$$ p = u^2, \quad q = v_{-10}^2, $$
and subsequently,  using the first integral (\ref{Z2-case2-fi}),  set 
$$ f^2u^2v^2= 1\quad \Longleftrightarrow \quad f\, =\, \frac{\pm 1}{u\, v}\, ,
$$
system (\ref{Z2-case2-BT}) can be written in a polynomial form as
$$\partial_x u = u (u_{10}^2 v^2-u^2 v_{-10}^2) \pm u^2 v, \quad \partial_x v =
v (u_{10}^2 v^2 - u^2 v_{-10}^2) \mp u v^2.$$
\end{enumerate}

\subsection{Derivation of discrete systems}

Now we consider the difference Lax pair 
\begin{equation} \label{sl2-ddLP}
\Psi_{10}\, =\, M(p, q_{10}, f;c_1, c_2) \, \Psi\, , \quad \Psi_{01}\, =\, M(p,
q_{01}, g;1, 1)\, \Psi\, ,  
\end{equation}
where matrix $M$ is given in (\ref{DT-sl2-gen}) and at least one of the
constants $c_1$,  $c_2$ is different from 0. It follows from the above system that
\begin{subequations} \label{sl2-res-eq}
\begin{eqnarray}
&& f\,   g_{10}\, -\, g\,   f_{01}  = 0, \label{consLaw} \\
&& f_{01} \, q_{11} - f\,  q_{10} - c_1 g_{10}\,  q_{11} + c_2 g \,  q_{01}=0, 
\\
&& f_{01}\,  p_{01} - f\, p - c_2 g_{10}\,  p_{10} + c_1 g\, p = 0, \\
&& f_{01}-f- c_1 (g_{10}-g) - f \, g_{10}\,  p_{10}\,  q_{10} + g \, f_{01}\, 
p_{01}\,  q_{01} = 0.
\end{eqnarray}
\end{subequations}
We can solve equations (\ref{sl2-res-eq}) for $p_{01}$,  $q_{01}$,  $f_{01}$ and
$g$ (or for $p_{10}$,  $q_{10}$,  $f$ and $g_{10}$). If $c_1=c_2=1$,  we derive
two sets of solutions,  as in the case of the nonlinear Schr{\"o}dinger.
Specifically,  the first branch is the singular solution already given in
(\ref{triv-sol}),  while the second branch involves rational expressions of the
remaining variables. When either $c_1$ or $c_2$ is equal to 0,  then system
(\ref{sl2-res-eq}) admits a unique non-trivial solution. This solution is given
by 
\begin{subequations} \label{SL2-sol}
\begin{eqnarray}
p_{01} &=&\frac{A}{f\, B^2}\, \left(f^2 p^2 q_{10} + c_2 f p (g_{10} p_{10}
q_{10}-1) - c_2^2 g_{10} p_{10} + c_1 c_2 g_{10} p \right),  \quad  f_{01}~=~f\,
 \frac{B}{A},   \\
q_{01} &=& \frac{B}{g_{10} A^2}\, \left(f (q_{11}-q_{10} + g_{10} p_{10} q_{10}
q_{11}) + c_1 g_{10} q_{11} (g_{10} p_{10} q_{11}-1)\right),  \, \, 
g~=~g_{10}\, \frac{A}{B}, 
\end{eqnarray}
\end{subequations}
where $A:=fp q_{11} + c_2 (g_{10} p_{10} q_{11}-1)$ and $B:= f p q_{10} + c_1
g_{10} p q_{11} - c_2$.

In this discrete context,  the first integrals of the B{\"a}cklund
transformation given in (\ref{sl2-D-con-det-gen}) become first integrals for
system (\ref{SL2-sol}),  i.e.
\begin{equation} \label{sl2-fi}
({\cal{T}}-1)\left(f^2 p\, q_{10}-c_2 f\right)=0, \quad ({\cal{S}}-1) \left(g^2
p\, q_{01}- g\right)=0.
\end{equation}
Moreover,  a generalised symmetry of the latter system follows from
(\ref{sl2-D-sym-gen}) and it is generated by the differential-difference
equations
\begin{eqnarray*}
&& \partial_x p = p\, \left(p_{10}\, q_{10}-p\, q \right) - \frac{c_2 p_{10}-
c_1 p}{f} =  p\, \left(p_{01}\, q_{01}-p\, q \right) - \frac{p_{01}-p}{g}, \\
&& \partial_x q =  q \left(p\, q-p_{-10}\, q_{-10} \right)-\frac{c_1 q- c_2
q_{-10}}{f_{-10}}=q \left(p\, q-p_{0, -1}\, q_{0, -1} \right)-\frac{q-q_{0,
-1}}{g_{0, -1}}, \\
&& \partial_x f =  f \left(p\, q-p_{10}\, q_{10} \right), \qquad \partial_x g = 
g \left(p\, q-p_{01}\, q_{01} \right).
\end{eqnarray*}

\subsubsection{First integrals and a seven point scalar difference equation}

Let us consider now system (\ref{sl2-res-eq}) with $c_1=c_2=1$ and try to
implement the first integrals (\ref{sl2-fi}) so that to reduce the number of
functions involved in this system by setting
\begin{equation} \label{sl2-fi-values}
f^2 p\, q_{10}\, -\,  f \, =\,  \alpha(n),  \qquad g^2 p\, q_{01}\, -\,  g\, =\,
\beta(m).
\end{equation}
One option is to use the above relations to replace $f$,  $g$ in terms of $p$
and $q$. In this case,  we must solve equations (\ref{sl2-fi-values}),  which
are quadratic in $f$ and $g$,  and hence introduce square roots,  and finally
derive a system of non-polynomial equations (correspondences) for $p$ and $q$. 

Another option is,  instead of eliminating $f$ and $g$,  to use relations
(\ref{sl2-fi-values}) to replace the shifts of $q$. In this case, equations (\ref{sl2-fi-values}) imply
\begin{equation} \label{sl2-fi-q-sub}
q_{10}\, =\, \frac{\alpha(n)+ f}{f^2 p}, \quad q_{01}\, =\,
\frac{\beta(m)+g}{g^2 p}.
\end{equation}
Moreover,  equation (\ref{consLaw}) suggests to introduce a potential $u$
through the relations
\begin{equation} \label{sl2-fg-sub}
f\, =\, \frac{u_{10}}{u}\, , \quad g\, =\, \frac{u_{01}}{u}.
\end{equation}
Additionally,  we introduce $v$ by $v := p/u$ for convenience.

Applying all the above substitutions to system (\ref{sl2-res-eq}),  we derive a
system for $v$ and $u$,  namely
\begin{equation}\label{sl2-red-sys-fi}
\frac{u_{11} + \alpha(n) u_{01}}{v_{01}}\, -\,  \frac{u_{11} + \beta(m)
u_{10}}{v_{10}}\, =\, 0\, , \quad u_{11} \left(v_{10}-v_{01}\right) \, +\, v
\left(u_{10}-u_{01}\right) \, =\, 0, 
\end{equation}
while a symmetry for this system is generated by
\begin{equation} \label{sl2-red-sys-fi-sym}
\partial_x u\, =\, \frac{-v}{v_{-10}}\left(u+\alpha(n-1) u_{-10}\right), \quad
\partial_x v\, =\, \frac{u}{u_{10}}\left(v+\alpha(n) v_{10}\right)\, =\,
\frac{u}{u_{01}}\left(v+\beta(m) v_{01}\right). 
\end{equation}

From equations (\ref{sl2-red-sys-fi}) we can derive a higher order scalar
equation either for $u$ or for $v$, namely 
\begin{equation} \label{sl2-red-sys-fi-u}
\left({\cal{S}}-1\right)\log X(n, u, u_{-10}) - \left({\cal{T}}-1\right)\log
Y(m, u, u_{0, -1}) + \left({\cal{S}}{\cal{T}}^{-1}-1\right) \log Z(n, m, u,
u_{-11}) \, =\,  0, 
\end{equation}
and
\begin{equation} \label{sl2-red-sys-fi-v}
\left({\cal{S}}-1\right)\log X(n, v_{-10}, v) - \left({\cal{T}}-1\right)\log
Y(m, v_{0, -1}, v) + \left({\cal{S}}{\cal{T}}^{-1}-1\right) \log Z(n, m,
v_{-11}, v) = 0,
\end{equation} 
where
\begin{equation} \label{sl2-red-XYZ}
X(n, u, x) = 1 + \alpha(n-1)\frac{x}{u}, \ Y(m, u, y) = 
1 + \beta(m-1)\frac{y}{u}, \ Z(n, m, u, z)= \frac{z-u}{\alpha(n-1) z - \beta(m) u}.
\end{equation}
A symmetry for equation (\ref{sl2-red-sys-fi-u}) follows from
(\ref{sl2-red-sys-fi-sym}) and it is generated by
$$ \partial_x u =  u \, X(n, u, u_{-10})\,  Y(m+1, u_{01}, u)\,  Z(n, m, u,
u_{-11})$$
while
$$ \partial_x v  = v \, X(n, v_{-10}, v)\,  Y(m+1, v, v_{01})\,  Z(n, m, v_{-11},
v) $$
generates a symmetry for equation (\ref{sl2-red-sys-fi-v}).

\begin{figure}[ht]
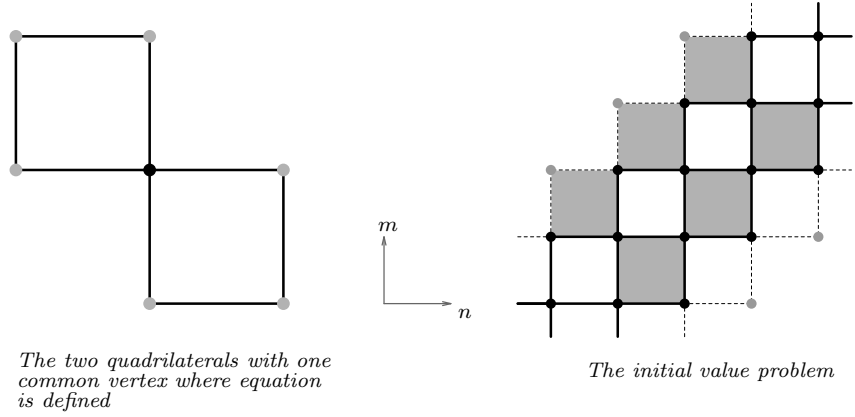

\centertexdraw{
\setunitscale 0.35
\move(-4.5 -2) \linewd 0.02 \setgray 0.4 \arrowheadtype t:V \arrowheadsize l:.12
w:.06 \avec(-4.5 -1) 
\move(-4.5 -2) \arrowheadtype t:V  \avec(-3.5 -2)
\setgray 1.0
\move (-2 -1) \lvec (-1 -1) \lvec (-1 0) \lvec (-2 0) \lvec (-2 -1) \lfill f:0.7
\move (-1 -2) \lvec (0 -2) \lvec (0 -1) \lvec (-1 -1) \lvec (-1 -2) \lfill f:0.7
\move (-1 0) \lvec (0 0) \lvec (0 1) \lvec (-1 1) \lvec (-1 0) \lfill f:0.7
\move (0 -1) \lvec (1 -1) \lvec (1 0) \lvec (0 0) \lvec (0 -1) \lfill f:0.7
\move (0 1) \lvec (1 1) \lvec (1 2) \lvec (0 2) \lvec (0 1) \lfill f:0.7
\move (1 0) \lvec (2 0) \lvec (2 1) \lvec (1 1) \lvec (1 0) \lfill f:0.7
\setgray 0.0
\linewd 0.04 \move (-8 -2) \lvec (-6 -2) \lvec (-6 0) \lvec (-10 0) \lvec (-10
2) \lvec(-8 2) \lvec (-8 -2)
\move (-8 -2) \fcir f:0.7 r:0.095 \move (-6 -2) \fcir f:0.7 r:0.095
\move (-6 0) \fcir f:0.7 r:0.095 \move (-8 0) \fcir f:0.0 r:0.095
\move (-10 0) \fcir f:0.7 r:0.095 \move (-10 2) \fcir f:0.7 r:0.095 \move (-8 2)
\fcir f:0.7 r:0.095
\move (-2 -2.5) \lvec (-2 -2) \lvec (-1 -2) \lvec (-1 -1) \lvec (0 -1) \lvec (0
0) \lvec (1 0) \lvec(1 1) \lvec (2 1) \lvec(2 2) \lvec(2.5 2)
\linewd 0.04  \move (-2 -2) \lvec (-2 -1) \lvec(-1 -1) \lvec (-1 0) \lvec (0 0)
\lvec (0 1) \lvec(1 1) \lvec (1 2) \lvec (2 2) \lvec (2 2.5) \move(-2.5 -2)
\lvec(-2 -2) \move(-2.5 -2) \lvec(-2 -2)
\move (-1 -2.5) \lvec (-1 -2) \lvec(0 -2) \lvec(0 -1) \lvec(1 -1) \lvec(1 0)
\lvec(2 0) \lvec(2 1) \lvec(2.5 1)
\linewd 0.001
\lpatt (.05 .05) \move (-2.5 -1) \lvec (-2 -1)  \lvec (-2 0) \lvec (-1 0) \lvec
(-1 1) \lvec (0 1) \lvec (0 2) \lvec (1 2) \lvec (1 2.5)
\move (0 -2.5) \lvec (0 -2) \lvec (1 -2) \lvec (1 -1) \lvec (2 -1) \lvec (2 0)
\lvec (2.5 0)
\lpatt  (1 1)
\move (-2 -2) \fcir f:0.0 r:0.075 \move (-2 -1) \fcir f:0.0 r:0.075
\move (-1 -2) \fcir f:0.0 r:0.075 \move (0 -2) \fcir f:0.0 r:0.075
\move (-1 -1) \fcir f:0.0 r:0.075 \move (-1 0) \fcir f:0.0 r:0.075 \move (0 -1)
\fcir f:0.0 r:0.075 \move (1 -1) \fcir f:0.0 r:0.075
\move (0 0) \fcir f:0.0 r:0.075 \move (0 1) \fcir f:0.0 r:0.075 \move (1 0)
\fcir f:0.0 r:0.075 \move (2 0) \fcir f:0.0 r:0.075 
\move (1 1) \fcir f:0.0 r:0.075 \move (2 1) \fcir f:0.0 r:0.075
\move (2 2) \fcir f:0.0 r:0.075 \move (1 2) \fcir f:0.0 r:0.075
\move (-2 0) \fcir f:0.6 r:0.075 
\move (-1 1) \fcir f:0.6 r:0.075 \move (0 2) \fcir f:0.6 r:0.075
\move (1 -2) \fcir f:0.6 r:0.075 \move (2 -1) \fcir f:0.6 r:0.075
\htext (-3.4 -2.2) {\scriptsize{$n$}}
\htext (-4.6 -.9) {\scriptsize{$m$}}
\htext (-10 -3) {\scriptsize{\it{The two quadrilaterals with one}}}
\htext (-10 -3.3) {\scriptsize{\it{common vertex where equation}}}
\htext (-10 -3.6) {\scriptsize{\it{is defined}}}
\htext (-1.5 -3.15) {\scriptsize{\it{The initial value problem}}}
}
\caption{{\small{\em{The stencil of seven points and the initial value
problem}}}} \label{fig-ivp-y}
\end{figure}

Equations (\ref{sl2-red-sys-fi-u}),  (\ref{sl2-red-sys-fi-v}) are similar and
have the same properties. They are defined on a stencil of seven points and can
be solved uniquely with respect to any $u_{ij}$ and $v_{ij}$ except $u$ and $v$,
 respectively. Because of this feature,  if initial data are given along a
double staircase,  then these equations uniquely determine the evolution above
and below this initial configuration as it is shown in Figure \ref{fig-ivp-y}.
\\

\noindent {\bf{Remark.}} When $\alpha(n)$,  $\beta(m)$ are constants,  i.e.
$\alpha(n)=\alpha$,  $\beta(m)=\beta$,  equations (\ref{sl2-red-sys-fi-u}) and
(\ref{sl2-red-sys-fi-v}) are related to the discrete Toda equation
\begin{equation} \label{Toda-H}
\left({\cal{S}}-1\right)\log({\rm{e}}^{w-w_{-10}}+1) +
\left({\cal{T}}-1\right)\log({\rm{e}}^{w_{-10}-w}-1) +
\left({\cal{S}}{\cal{T}}-1\right) \log
\frac{{\rm{e}}^{w-w_{-1-1}}+\gamma}{{\rm{e}}^{w-w_{-1-1}}+1}\, =\,  0,  \quad
\gamma\, :=\, \frac{\alpha}{\beta}\, \, , 
\end{equation}
i.e. equation (H) in \cite{A-JNMP}. This relation is made evident if we first
reverse the $m$ direction,  i.e. change indices $(i, j)$ to $(i, -j)$ and
operator $\cal{T}$ to its inverse ${\cal{T}}^{-1}$ in both equations
(\ref{sl2-red-sys-fi-u}),  (\ref{sl2-red-sys-fi-v}),  and then make the point
transformation
$$u\, =\, (-1)^m \alpha^n \beta^{m}{\rm{e}}^{-w}\quad {\mbox{and}} \quad v\, =\,
(-1)^m \alpha^{-n} \beta^{-m}{\rm{e}}^{w}, $$
to each equation,  respectively. In this context,  system (\ref{sl2-red-sys-fi})
defines the self-duality transformation for the Toda equation (\ref{Toda-H})
\cite{A-JNMP}. In particular,  if we make the above change of variables to
system (\ref{sl2-red-sys-fi}),  then it will become
\begin{equation}\label{autoBT-Toda-H}
{\rm{e}}^{\tilde{w}_{10}-\tilde{w}}\, =\, \frac{{\rm{e}}^{w_{10}-w_{0, -1}}+
\gamma }{{\rm{e}}^{w_{10}-w_{0, -1}}+1} \left({\rm{e}}^{w_{1,
-1}-w_{10}}-1\right), \quad {\rm{e}}^{\tilde{w}_{0, -1}-\tilde{w}}\, =\,
\frac{1}{\gamma}\, \frac{{\rm{e}}^{w_{10}-w_{0, -1}}+ \gamma
}{{\rm{e}}^{w_{10}-w_{0, -1}}+1} \left({\rm{e}}^{w_{1, -1}-w_{0, -1}}+1\right), 
\end{equation}
where $w$ and $\tilde{w}$ are two different solutions of equation
(\ref{Toda-H}).

\subsubsection{Lax pair with matrix (\ref{Z2-case1-DM-M}) and a six point difference equation}

Let us consider now the Lax pair
$$\Psi_{10} = M(p) \Psi, \quad \Psi_{01} = M(p, q_{01}, g;1, 1) \Psi, $$
where matrix $M(p)$ is given in (\ref{Z2-case1-DM-M}) and $M(p, q_{01}, g;1, 1)$ in
(\ref{DT-sl2-gen}). It follows from the compatibility condition of the above
pair that
$$ q_{10} = p, \quad g\, =\, \frac{p_{01}-p}{p (p_{01} p_{-11}-p p_{10})}, \quad
g_{10}\, =\, \frac{p}{p_{01}} g, $$
and finally we  arrive at the six point  difference equation
\begin{equation} \label{Z2-scalar-eq}
\frac{p_{11}-p_{10}}{p_{10} (p_{11} p_{01}-p_{10} p_{20})} =
\frac{p_{01}-p}{p_{01} (p_{01} p_{-11}-p p_{10})}. 
\end{equation}
We also find a first integral and a symmetry of this equation,  which are given
by
$$({\cal{S}}-1)\frac{(p-p_{01}) (p_{10}-p_{-11})}{(p_{01} p_{-11}-p p_{10})^2} =
0 \quad {\mbox{and}} \quad \partial_x p\, =\, p^2 \left(p_{10}-p_{-10}\right), 
$$
respectively.

\begin{figure}[ht]
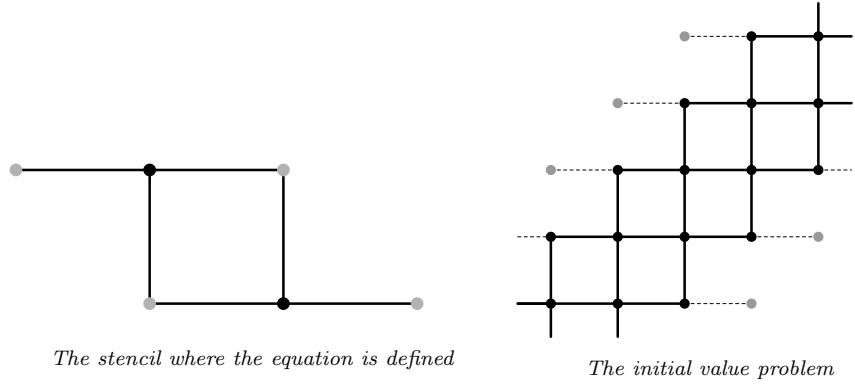

\centertexdraw{
\setunitscale 0.35
\setgray 0.0
\linewd 0.04 \move (-8 -2) \lvec (-6 -2) \lvec (-6 0) \lvec (-10 0) \move (-8 0)
\lvec (-8 -2) \move (-6 -2) \lvec (-4 -2)
\move (-8 -2) \fcir f:0.7 r:0.095 \move (-6 -2) \fcir f:0.0 r:0.095
\move (-6 0) \fcir f:0.7 r:0.095 \move (-8 0) \fcir f:0.0 r:0.095
\move (-10 0) \fcir f:0.7 r:0.095 \move (-4 -2) \fcir f:0.7 r:0.095
\move (-2 -2.5) \lvec (-2 -2) \lvec (-1 -2) \lvec (-1 -1) \lvec (0 -1) \lvec (0
0) \lvec (1 0) \lvec(1 1) \lvec (2 1) \lvec(2 2) \lvec(2.5 2)
\linewd 0.04  \move (-2 -2) \lvec (-2 -1) \lvec(-1 -1) \lvec (-1 0) \lvec (0 0)
\lvec (0 1) \lvec(1 1) \lvec (1 2) \lvec (2 2) \lvec (2 2.5) \move(-2.5 -2)
\lvec(-2 -2) \move(-2.5 -2) \lvec(-2 -2)
\move (-1 -2.5) \lvec (-1 -2) \lvec(0 -2) \lvec(0 -1) \lvec(1 -1) \lvec(1 0)
\lvec(2 0) \lvec(2 1) \lvec(2.5 1)
\linewd 0.001
\lpatt (.05 .05) \move (-2 0) \lvec (-1 0) \move (0 1) \lvec (-1 1) \move (0 2)
\lvec (1 2) \move (-2.5 -1) \lvec (-2 -1) 
\move  (0 -2) \lvec (1 -2) \move (1 -1) \lvec (2 -1) \move (2 0) \lvec (2.5 0)
\lpatt  (1 1)
\move (-2 -2) \fcir f:0.0 r:0.075 \move (-2 -1) \fcir f:0.0 r:0.075
\move (-1 -2) \fcir f:0.0 r:0.075 \move (0 -2) \fcir f:0.0 r:0.075
\move (-1 -1) \fcir f:0.0 r:0.075 \move (-1 0) \fcir f:0.0 r:0.075 \move (0 -1)
\fcir f:0.0 r:0.075 \move (1 -1) \fcir f:0.0 r:0.075
\move (0 0) \fcir f:0.0 r:0.075 \move (0 1) \fcir f:0.0 r:0.075 \move (1 0)
\fcir f:0.0 r:0.075 \move (2 0) \fcir f:0.0 r:0.075 
\move (1 1) \fcir f:0.0 r:0.075 \move (2 1) \fcir f:0.0 r:0.075
\move (2 2) \fcir f:0.0 r:0.075 \move (1 2) \fcir f:0.0 r:0.075
\move (-2 0) \fcir f:0.6 r:0.075 
\move (-1 1) \fcir f:0.6 r:0.075 \move (0 2) \fcir f:0.6 r:0.075
\move (1 -2) \fcir f:0.6 r:0.075 \move (2 -1) \fcir f:0.6 r:0.075
\htext (-9.5 -3) {\scriptsize{\it{The stencil where the equation is defined}}}
\htext (-1.5 -3.15) {\scriptsize{\it{The initial value problem}}}
}
\caption{{\small{\em{The stencil of six points and the initial value problem for
equation (\ref{Z2-scalar-eq})}}}} \label{fig-ivp-Z2-p}
\end{figure}

It is worth noting that equation (\ref{Z2-scalar-eq}) can be uniquely solved
with respect to any value of $p$ except $p_{10}$ and $p_{01}$. If initial data
are given along a double staircase as it is shown in Figure \ref{fig-ivp-Z2-p}, 
which must be consistent with the first integral,  then the evolution of these
data is uniquely determined above and below the double staircase by equation
(\ref{Z2-scalar-eq}).\\

\noindent {\bf{Remark.}} If we set the value of the above first integral to
$\alpha(m)$ and,  subsequently,   make the change of independent variables $(n,
m) \mapsto (k, l) := (n+m, m)$,  then we will arrive at the following
quadrilateral equation for $\tilde{p}(k, l) = p(n, m)$.
$$\left(\tilde{p}-\tilde{p}_{11}\right)\,
\left(\tilde{p}_{10}-\tilde{p}_{01}\right)\, =\, \alpha(l)\, \left(\tilde{p}
\tilde{p}_{10}-\tilde{p}_{01} \tilde{p}_{11}\right)^2.$$

\section{$\field{Z}_2$ reduction group: Generic orbit} \label{sec-Z2-gen}

A $\field{Z}_2$ invariant Lax operator with simple poles in the generic 
orbit can be  taken in the form 
\begin{equation}\label{Lz2gen}
{\cal{L}}\, =\, D_x+\frac{1}{\lambda-1}S\, \, -\, \frac{1}{\lambda+1}\sigma_3 S
 \sigma_3, \qquad S := \frac{1}{p-q}\, \left(\begin{array}{cc} p+q &
-2\, p\, q \\2 & -p-q \end{array}\right).
\end{equation}
The corresponding NLS type equation is
$$
p_t \ = \ p_{xx}-\frac{2p_x^2}{p-q}+\frac{8pqp_x-4p^2q_x}{(p-q)^2}\,,\quad 
q_t \ = \ -q_{xx}-\frac{2q_x^2}{p-q}+\frac{8qpq_x-4q^2p_x}{(p-q)^2}\,,
$$
which is actually equation (m) in \cite{mshy}.

The Darboux matrix for the above Lax operator is derived in the same way as in
the previous section and three distinct cases occur.

\begin{enumerate}
\item The first  Darboux matrix is
\begin{equation}
M\, =\, \lambda \left(\begin{array}{cc} 0 & a_1 \\ a_2 & 0 \end{array}\right) +
\left(\begin{array}{cc} b_1 u & 0 \\ 0 & b_2 v \end{array} \right),  \quad u v =
1,  \quad a_i,  b_i \in {\mathbb{C}}\quad {\mbox{and}} \quad |a_1 a_2|^2 + |b_1
b_2|^2 \ne 0, 
\end{equation}
and the B{\"a}cklund transformation is given by
\begin{equation}
p_{10} = \frac{a_1+ b_1 p \, u}{b_2 v+ a_2 p},  \quad q_{10} = \frac{a_1+ b_1
q\, u}{b_2 v+ a_2 q}, \quad \partial_x u = 4\, u\,
\left(\frac{p_{10}}{p_{10}-q_{10}}-\frac{p}{p-q} \right).
\end{equation}
This transformation contains as particular subcases two Darboux transformations
related to point symmetries,  namely scalings ($a_1=a_2=0$,  $u=v=1$) and
inversions  ($b_1=b_2=0$,  $a_1=a_2=1$).

\item The second  Darboux matrix is
\begin{equation} \label{Z2-M-D1}
M(p, q_{10})\, =\, \frac{1}{\lambda-1} \left(\begin{array}{cc} q_{10} & 1\\1
& -p \end{array} \right) - \frac{1}{\lambda+1}\left(\begin{array}{cc} q_{10} &
-1\\-1 & -p \end{array} \right), \end{equation}
and the B{\"a}cklund transformation is given by
\begin{equation}
q_{10} = \frac{-1}{p}, \quad \partial_x p\, =\, 4\, p\, \left(\frac{p}{p-q}\,
-\, \frac{p_{10}}{p_{10}-q_{10}} \right).
\end{equation}

\item The last  Darboux matrix is given by
\begin{equation} \label{Z2-M-gen}
M(p, q_{10}, f;c_1, c_2)\, =\, \frac{f}{\lambda-1} \left(\begin{array}{cc}
q_{10} & -p\, q_{10} \\ 1 & -p \end{array} \right) -
\frac{f}{\lambda+1}\left(\begin{array}{cc} q_{10} & p\, q_{10} \\ -1 & -p
\end{array} \right) + \left(\begin{array}{cc} c_1 &0 \\ 0 & c_2 \end{array}
\right), 
\end{equation}
where $c_1$,  $c_2$ are constants such that $|c_1|^2+|c_2|^2 \ne 0$ and, 
without loss of generality,  we can set these constants equal to 0 or 1. The
derivatives of $p$,  $q_{10}$ and $f$ are given by the following relations
\begin{eqnarray}
\partial_x p &=& 4\, p\, \left(\frac{p_{10}}{q_{10}-p_{10}} -
\frac{p}{q-p}\right) + \frac{2}{f} \, \frac{c_2 p_{10}-c_1 p}{q_{10}-p_{10}},
\nonumber \\
\partial_x q_{10} &=& -4\, q_{10}\, \left(\frac{p_{10}}{q_{10}-p_{10}} -
\frac{p}{q-p}\right) + \frac{2}{f}\, \frac{c_2 q_{10}-c_1 q}{q-p}\, , 
\label{z2-sym} \\ 
\partial_x f &=& \frac{2 c_1}{q_{10}-p_{10}}\, -\, \frac{2 c_2}{q-p}. \nonumber
\end{eqnarray}
Function
\begin{equation} \label{z2-fi}
\Phi(c_1, c_2) = \left(2\, f\, p+c_2\right) \left(2\, f\, q_{10}-c_1\right)
\end{equation}
defines a first integral for equations (\ref{z2-sym}),  i.e. $D_x \Phi(c_1, c_2)
=0$ on solutions of the latter system.
\end{enumerate}

\subsection{Derivation of discrete systems}

The first discrete Lax pair to consider is 
\begin{equation} \label{z2-ddLP}
\Psi_{10}\, =\, M(p, q_{10}, f;c_1, c_2) \, \Psi\, , \quad \Psi_{01}\, =\, M(p,
q_{01}, g;1, 1)\, \Psi\, ,  
\end{equation}
where matrix $M$ is given in (\ref{Z2-M-gen}) and for constants $c_1$,  $c_2$
one may consider three distinct cases : ({\rm{i}}) $c_1=c_2=1$,  ({\rm{ii}})
$c_1=1$,  $c_2=0$ and ({\rm{iii}}) $c_1=0$,  $c_2=1$.

In this generic setting,  the compatibility condition of system (\ref{z2-ddLP})
results to
\begin{subequations} \label{z2-gen-ddsys}
\begin{eqnarray}
&& f_{01} - f - c_1 g_{10} + c_2 g = 0, \\
&& f_{01}\, p_{01}\, q_{11} - f\, p\, q_{10}- c_2 g_{10}\, p_{10}\, q_{11} + c_1
g\, p\, q_{01} = 0 , \\
&& f_{01} q_{11} - f q_{10}  - c_1 g_{10} q_{11} + c_1 g q_{01} \, -\,  2\, 
q_{11} (g f_{01} q_{01} - g_{10} f q_{10})=0,  \\
&&  f_{01}\, p_{01} - f\, p - c_2 g_{10}\, p_{10} + c_2 g\, p \, +\,  2\, p\, 
(g\, f_{01}p_{01}\, -\, g_{10}\, f\, p_{10}) = 0.
\end{eqnarray}
\end{subequations}
This system can be solved for either $(p_{01}, q_{01}, f_{01}, g)$ or $(p_{10},
q_{10}, f, g_{10})$. When $c_1=c_2=1$,  then it leads to a solution with two
branches: one branch is the trivial solution (\ref{triv-sol}),  while the
non-trivial branch involves rational expressions of the remaining variables. In
the other two cases ($c_1=1$,  $c_2=0$ or $c_1=0$,  $c_2=1$),  system
(\ref{z2-gen-ddsys}) admits a unique non-trivial solution. In any case,  the
non-trivial branch can be easily found,  but is omitted here because of its
length,  and we consider it as a difference system. For this system,  it can be
verified that it admits two first integrals, 
\begin{equation} \label{z2-gen-fi}
({\cal{T}}-1)\left(2\, f\, p+c_2\right) \left(2\, f\, q_{10}-c_1\right)=0, \quad
 ({\cal{S}}-1)\left(2\, g\, p+1\right) \left(2\, g\, q_{01}-1\right)=0, 
\end{equation}
and a symmetry generated by
\begin{eqnarray}
&&\partial_x p = 2\, p\, \left(\frac{p_{10}}{q_{10}-p_{10}} -
\frac{p}{q-p}\right) + \frac{1}{f} \, \frac{ c_2 p_{10}- c_1 p}{q_{10}-p_{10}}  
=  2\, p\, \left(\frac{p_{01}}{q_{01}-p_{01}} - \frac{p}{q-p}\right) +
\frac{1}{g} \, \frac{p_{01}- p}{q_{01}-p_{01}} ,  \nonumber\\
&& \partial_x q = -2\, q\, \left(\frac{p}{q-p} -
\frac{p_{-10}}{q_{-10}-p_{-10}}\right) + \frac{1}{f_{-10}}\, \frac{c_2 q-c_1
q_{-10}}{q_{-10}-p_{-10}} \label{z2-gen-sym} \\
 && {\phantom{\partial_x q}} = -2\, q\, \left(\frac{p}{q-p} - \frac{p_{0,
-1}}{q_{0, -1}-p_{0, -1}}\right) + \frac{1}{g_{0, -1}}\, \frac{q- q_{0,
-1}}{q_{0, -1}-p_{-10}}, \nonumber\\
&&\partial_x f = \frac{c_1}{q_{10}-p_{10}}\, -\, \frac{c_2}{q-p}, \quad
\partial_x g = \frac{1}{q_{01}-p_{01}}\, -\, \frac{1}{q-p}.\nonumber
\end{eqnarray}

We can use the two first integrals (\ref{z2-gen-fi}) to reduce the number of
dependent variables involved in system (\ref{z2-gen-ddsys}). In particular,  we
have two different options. The first option is to use the first integrals to
remove function $q$ from the system and a conservation law to replace $f$ and
$g$ with a potential $u$,  as we  did in the previous section. The second option
is to consider particular values for these integrals so that to eliminate $f$
and $g$. These considerations are presented in the following two subsections.

\subsubsection{First integrals and a seven point scalar equation}

Let us consider the case $c_1=c_2=1$ for system (\ref{z2-gen-ddsys}) and its
integrals (\ref{z2-gen-fi}). Choosing the values of the latter, 
\begin{equation} \label{z2-red-fi-1}
\left(2\, f\, p+1\right) \left(2\, f\, q_{10}-1\right)\, =\, \alpha(n)-1, \qquad
\left(2\, g\, p+1\right) \left(2\, g\, q_{01}-1\right)\, =\, \beta(m)-1, 
\end{equation}
we can express $q_{10}$ and $q_{01}$ in terms of $p$,  $f$ and $g$ as
\begin{equation} \label{z2-red-q-sub}
q_{10} \, =\, \frac{1}{2 f}\, \frac{2fp+\alpha(n)}{2 f p +1}\, , \quad q_{01} \,
=\, \frac{1}{2 g}\, \frac{2gp+\beta(m)}{2 g p +1}\, .
\end{equation}
Moreover,  the first equation of (\ref{z2-gen-ddsys}) for $c_1=c_2=1$ has the
form of a conservation law,  suggesting the introduction of a potential $u$ via
the relations
\begin{equation} \label{z2-red-fg-u}
f\, =\, u_{10}-u\, , \quad g\, =\, u_{01}-u\, .
\end{equation}
We use now relations (\ref{z2-red-q-sub}),  (\ref{z2-red-fg-u}) to eliminate
$q$,  $f$ and $g$ from system (\ref{z2-gen-ddsys}) and derive the following
system for $p$ and $u$.
\begin{subequations}  \label{z2-red-up}
\begin{eqnarray}
&& 2\, p_{10}\, =\,  \frac{\alpha(n)-\beta(m)}{u_{10}-u_{01}}\, -\,
\frac{\beta(m)}{u_{11}-u_{10}}\, -\, \frac{2 (\alpha(n)-1) p}{1+ 2 p
(u_{10}-u)}\, , \\
&& 2\,  p_{01} \, =\, \frac{\alpha(n)-\beta(m)}{u_{10}-u_{01}}\, -\,
\frac{\alpha(n)}{u_{11}-u_{01}}\, -\, \frac{2 (\beta(m)-1) p}{1 + 2 p
(u_{01}-u)}\, .
\end{eqnarray} 
\end{subequations}
A symmetry of this system easily follows from (\ref{z2-gen-sym}) by using
substitutions (\ref{z2-red-q-sub}),  (\ref{z2-red-fg-u}) but it is omitted here
because of its length. Equations (\ref{z2-red-up}) can be solved uniquely either
for the pair $(p_{10}, u_{10})$ or for $(p_{01}, u_{01})$,  but here we present
it in this form because it is more elegant and convenient. Moreover it makes
apparent the invariance of the system under the involution $(p_{ij}, u_{ij},
\alpha(n), \beta(m)) \leftrightarrow (p_{ji}, u_{ji}, \beta(m), \alpha(n))$.
Regarding the Cauchy problem,  initial values along a staircase are compatible
with the evolution defined by the above system.

Equations (\ref{z2-red-up}) can be decoupled to a scalar equation for $u$.
Indeed,  the compatibility condition ${\cal{T}}\left(p_{10}\right) = {\cal{S}}
\left( p_{01}\right)$ implies that $u$ must obey the equation
\begin{equation} \label{z2-red-u-eq}
\left({\cal{S}}-1\right) \frac{\alpha(n-1)}{u-u_{-10}}\, -\, \left(
{\cal{T}}-1\right) \frac{\beta(m-1)}{u-u_{0, -1}}\, +\, \left( {\cal{S}}
{\cal{T}}^{-1}-1 \right)\frac{\beta(m)-\alpha(n-1)}{u-u_{-11}}\, =\, 0,
\end{equation}
which, up to point transformations, is the non-autonomous version of the Toda-type equation (A) in
\cite{A-JNMP},  cf. also \cite{A-JPA}. A symmetry of this equations follows from the symmetry of system
(\ref{z2-red-up}) and is generated by
\begin{equation} \label{z2-red-u-sym}
\partial_x u\, =\, \frac{(u_{10}-u) (u_{0, -1}-u) (u_{1, -1}-u)}{F_{0, -1}}\,
=\, -\, \frac{(u_{-10}-u) (u_{01}-u) (u_{-11}-u)}{F_{-10}},  
\end{equation}
where
\begin{equation}\label{z2-red-R}
F_{00} := \alpha(n) (u-u_{01}) (u_{10}-u_{11}) \, -\, \beta(m) (u-u_{10})
(u_{01}-u_{11})\, .
\end{equation}
Equation (\ref{z2-red-u-eq}) is defined on a stencil of seven points and can be
solved uniquely with respect to any $u_{ij}$ except $u$. Because of this
property,  if initial data are given along a double staircase,  then equation
(\ref{z2-red-u-eq}) uniquely determine the evolution above and below this
initial configuration as it is shown in Figure \ref{fig-ivp-y}. \\

\noindent {\bf{Remark.}} Equation (\ref{z2-red-u-eq}) is the Euler-Lagrange
equation for the Lagrangian
$${\cal{L}}\,=\, \alpha(n-1) \log(u - u_{-10}) - \beta(m-1) \log (u-u_{0, -1})
-\left(\alpha(n-1)-\beta(m-1) \right) \log(u_{-10}-u_{0, -1}), $$
which is also considered as a Lagrangian for the discrete Schwarzian KdV or
Q1$_0$ \cite{NL},  the form of which is $F_{00} = 0$ \cite{NC95,  ABS},  where
$F$ is given in (\ref{z2-red-R}).

\subsubsection{First integrals and a five point scalar equation}

Now we consider the case $c_1=c_2=1$ and two particular values for the
first integrals given in (\ref{z2-gen-fi}). More precisely,  let us consider
that
\begin{equation} \label{z2-gen-fi-val}
(2 f p+1) (2 f q_{10}-1) =0, \quad (2 g p +1) (2 g q_{01}-1) = -1, 
\end{equation}
from which we can express $f$ and $g$ in terms of $p$ and $q$ rationally. While
the second equation determines $g$ uniquely\footnote{The solution $g=0$ is not
considered since along it system (\ref{z2-gen-ddsys}) and its symmetry
(\ref{z2-gen-sym}) degenerate.},  the first equation admits two different
solutions and we choose\footnote{The second choice for $f$ leads to a system related to (\ref{z2-gen-red-1}) by a point transformation.}
\begin{equation} \label{z2-gen-fg-subs}
f\, =\, \frac{-1}{2 p}, \quad g\, =\,  \frac{1}{2}\left(\frac{1}{q_{01}}\, -\,
\frac{1}{p} \right).
\end{equation}
Then,  for $c_1=c_2=1$ and in view of substitutions (\ref{z2-gen-fg-subs}), 
system (\ref{z2-gen-ddsys}) and its symmetry (\ref{z2-gen-sym}) reduce to
\begin{subequations} \label{z2-gen-red-1}
\begin{equation}
p_{10}-p = q_{10}-q_{01}, \qquad \frac{1}{p_{10}}-\frac{1}{p_{01}} =
\frac{1}{q_{11}} - \frac{1}{q_{01}}
\end{equation}
and
\begin{equation}
\partial_x p = p^2 \left(\frac{1}{q_{10}-p_{10}}- \frac{1}{q-p}\right), \quad
\partial_x q = \frac{p q}{p-q}-\frac{p_{-10} q_{-10}}{p_{-10}-q_{-10}}, 
\end{equation}
respectively.
\end{subequations}

It can be readily verified that the above discrete system for $p$ and $q$ can be
written in a conserved form as
$$({\cal{S}}-1) (q-p) = ({\cal{T}}-1) q, \qquad  ({\cal{S}}-1)
\left(\frac{1}{p}-\frac{1}{q_{01}}\right) = ({\cal{T}}-1) \frac{1}{p}.$$
We can use either of these conserved forms to introduce a potential and then
derive an equation only for the potential. In either of the cases,  we end up
actually with the same scalar equation. Here,  we introduce potential $w$
employing the first conservation law and,  in particular,  we set 
$$ p = w_{0, -1}-w_{-1, 0}, \quad q = w_{0, -1}-w_{-1, -1}.$$
The substitution of the above expressions into equations (\ref{z2-gen-red-1})
results to a scalar equation for potential $w$,\footnote{If we used the
second conservation law to introduce the potential,  then the resulting equation
would be related to (\ref{z2-pot-eq}) by interchanging $n$ and $m$,  i.e.
changing indices $(ij)$ to $(ji)$.}
\begin{equation} \label{z2-pot-eq}
\frac{1}{w-w_{10}} + \frac{1}{w-w_{-10}} =  \frac{1}{w-w_{1,
-1}}+\frac{1}{w-w_{-11}}\,,
\end{equation}
and a symmetry of this equation is generated by
$$\partial_x w \, =\,  \frac{(w-w_{-10}) (w-w_{-11})}{w_{-10}-w_{-11}}. $$
A staircase initial value problem for equation (\ref{z2-pot-eq}) is similar to
the one we considered for the Toda equation in the previous section. That is, 
initial data can be given at points $w_{i, -i}$ and $w_{i, -i-1}$ from which a
solution can be uniquely determined on the whole lattice.\\

\noindent {\bf{Remark.}} By the change of independent variables $(n, m) \mapsto
(k, l) := (n+m, m)$,  equation (\ref{z2-pot-eq}) can be written as ``the missing
identity of Frobenius'' for the function $\tilde{w}(k, l) = w(n, m)$, 
$$\frac{1}{\tilde{w}-\tilde{w}_{10}} + \frac{1}{\tilde{w}-\tilde{w}_{-10}} = 
\frac{1}{\tilde{w}-\tilde{w}_{01}}+\frac{1}{\tilde{w}-\tilde{w}_{0, -1}}\, , $$
which appears in the theory of Pad{\'e} approximants \cite{Gragg},  as well as
in relation with the discrete KdV equations H1,  H3 \cite{NC95,  ABS} and the
$\epsilon$-algorithm \cite{PGR}.

\subsubsection{A Lax pair with matrix (\ref{Z2-M-D1}) and a six point difference equation}

Now we consider the discrete Lax pair
$$\Psi_{10} = M(p, q_{10}) \Psi, \quad \Psi_{01} = M(p, q_{01}, g;1, 1) \Psi, 
$$
where $M(p,q_{10})$ is given in (\ref{Z2-M-D1}) and $M(p, q_{01}, g;1, 1)$ in (\ref{Z2-M-gen}). The
compatibility condition of this system implies
$$ q_{10}\, =\, \frac{-1}{p}, \quad g\, =\, \frac{p_{-11}(p_{01}-p) (1+p
p_{10})}{2 p (p p_{10}-p_{01} p_{-11})}, \quad g_{10}\, =\, \frac{(p_{01}-p)
(1+p_{01} p_{-11})}{2 (p p_{10} - p_{01} p_{-11})}, $$
which subsequently leads to the scalar difference equation
\begin{equation} \label{Z2-red-scalar}
p_{10} p_{01} \Big\{(p_{11}+p_{-11}) (p p_{10} p_{20} + p_{01}) - (p+p_{20})
(p_{-11} p_{01} p_{11} + p_{10}) \Big\} + (1-p_{10} p_{01}) \Big(p p_{10}^2
p_{20} - p_{-11} p_{01}^2 p_{11} \Big)\, =\, 0.
\end{equation}
This equation admits the first integral
\begin{equation} \label{Z2-red-scalar-fi}
\Phi := \frac{pp_{10} p_{01} p_{-11}}{(p p_{10}-p_{01}p_{-11})^2} \left(p + \frac{1}{p_{10}} -
p_{01} - \frac{1}{p_{-11}} \right)
 \left(\frac{1}{p} + p_{10} -
\frac{1}{p_{01}} - p_{-11}\right).
\end{equation}
Moreover,  a generalised symmetry of (\ref{Z2-red-scalar}) is generated  by
$$\partial_x p\, =\, p\, \left(\frac{1}{1+p p_{10}} - \frac{1}{1+p p_{-10}}
\right). $$
Finally,  it can be easily shown that
a non-autonomous symmetry of equation (\ref{Z2-red-scalar}) is generated by
$$\partial_\tau p \, =\,  \left(\frac{n}{1+p p_{10}} - \frac{n-1}{1+p p_{-10}} -
\frac{1}{2} \right)\, p.$$

Equation (\ref{Z2-red-scalar}) is defined on a stencil of six points,  cf.
Figure \ref{fig-ivp-Z2-p},  and can be uniquely solved with respect to any value
of $p$ except $p_{10}$ and $p_{01}$. Initial data for equation
(\ref{Z2-red-scalar}) can be given along a double staircase as it is shown in
Figure \ref{fig-ivp-Z2-p}.\\

\noindent {\bf{Remark.}} If we set the value of the first integral
(\ref{Z2-red-scalar-fi}) to $\alpha(m)$ and,  subsequently,   make the change of
independent variables $(n, m) \mapsto (k, l) := (n+m, m)$,  then equation $\Phi
= \alpha(m)$ will become a quadrilateral equation (correspondence) for
$\tilde{p}(k, l) = p(n, m)$,  namely
\begin{equation} \label{Z2-Hirota}
\tilde{p}\tilde{p}_{10} \tilde{p}_{01} \tilde{p}_{11} H(\tilde{p})
H\left(\tilde{p}^{-1}\right) = \alpha(l) (\tilde{p} \tilde{p}_{10} -
\tilde{p}_{01} \tilde{p}_{11})^2, \qquad H(\tilde{p}) := \tilde{p} +
\frac{1}{\tilde{p}_{10}} - \frac{1}{\tilde{p}_{01}} - \tilde{p}_{11}. 
\end{equation}
Obviously if we set $\alpha(l) = 0$,  the above equation reduces to Hirota's
discrete KdV equation \cite{Hir} either in the form $H(\tilde{p})=0$ or
$H(\tilde{p}^{-1})=0$. Hence,  we consider equation (\ref{Z2-Hirota}) as a
quadratic Hirota KdV equation.  This relation allowed us to derive the
non-autonomous symmetry of equation (\ref{Z2-red-scalar}) from the corresponding
symmetries of Hirota's KdV equation \cite{MX}.

\section{Dihedral reduction group: Degenerate orbit} \label{sec-Dih}

We now consider Lax operators which are invariant with respect to the following
transformations
\begin{equation}  \label{reduct_group}
 s_1(\lambda):{\cal{L}}(\lambda) \rightarrow \sigma_3 {\cal{L}}(-\lambda)\sigma_3,
\quad s_2(\lambda):{\cal{L}}(\lambda) \rightarrow
\sigma_1 {\cal{L}}\left(\lambda^{-1}\right)\sigma_1, \quad
\sigma_1=\left(\begin{array}{cc} 0 & 1 \\ 1 & 0 \end{array} \right).
\end{equation}
Here,  the reduction group is generated by the above set of involutions and it
is isomorphic to $\field{Z}_2 \times \field{Z}_2\cong \field{D}_2$. The invariant Lax operator corresponding
to the degenerate orbit can be taken in the form
\begin{equation} \label{Dih-Lax-Op}
{\cal{L}}\ = \ D_x+\, \lambda^{2}\, \sigma_3\, +\lambda\,  \left(\begin{array}{cc} 0 & 2\,
p\\ 2\, q & 0\end{array}\right) \, +\, \frac{1}{\lambda}\, 
\left(\begin{array}{cc} 0 & 2\, q\\ 2\, p & 0\end{array}\right)\, -\,
\frac{1}{\lambda^2}\, \sigma_3.
\end{equation}
This operator corresponds to the following deformation of the derivative NLS equation \cite{mshy}
\begin{equation} \label{Dih-cont-sys}
p_t=p_{xx}+8\, (p^2\, q)_x-4\, q_x, ~~~q_t=-q_{xx}+8\, (p\, q^2)_x-4\, p_x. 
\end{equation}

It is simple to check that matrix $\sigma_3$ is a Darboux matrix for
operator (\ref{Dih-Lax-Op}) and corresponds to the discrete symmetry $(p, q)
\mapsto (-p, -q)$ of system (\ref{Dih-cont-sys}). A
$\lambda$-dependent Darboux matrix for operator (\ref{Dih-Lax-Op}) is 
\begin{equation}  \label{Dih-M}
M(p, q_{10}, f;u) = u\, \left(\left(\begin{array}{cc} \lambda^2&0 \\0 &
\lambda^{-2} \end{array}\right) + \lambda  \left(\begin{array}{cc} 0 & p
\\q_{10} & 0 \end{array}\right) + f\, \left(\begin{array}{cc} 1&0 \\0 &1
\end{array}\right) + \frac{1}{\lambda}  \left(\begin{array}{cc} 0& q_{10} \\ p
&0 \end{array}\right)\right)
\end{equation}
and the corresponding B{\"a}cklund transformation is given by
\begin{subequations}\label{Dih-D-cond}
\begin{eqnarray} 
\partial_x p &=& 2\, \Big((p_{10} q_{10}-p\, q) p + (p-p_{10}) f + q- q_{10}
\Big),  \\
\partial_x q_{10} &=& 2\, \Big((p_{10} q_{10}-p\, q) q_{10}  + p- p_{10} +
(q-q_{10}) f \Big), \\
\partial_x f &=& 2 \Big((p_{10} q_{10}-p\, q) f + (p-p_{10})p +  (q-q_{10})
q_{10} \Big), \\
\partial_x u &=& -2 (p_{10} q_{10}-p\, q) u.
\end{eqnarray}
\end{subequations}
It is straightforward to show that these differential equations admit two first
integrals $\partial_x \Phi^{(i)} = 0$,  $i=1, 2$,  where
\begin{equation} \label{dih-D-fi}
\Phi^{(1)} = u^2 \left( f- p\, q_{10} \right),  \quad \Phi^{(2)} = u^2 \left(f^2+1 -
p^2-q_{10}^2\right), 
\end{equation}
which imply that matrix $M$ has constant determinant since
$$\det M = \left(\lambda^2 + \frac{1}{\lambda^2} \right) \Phi^{(1)} + \Phi^{(2)}.$$

\subsection{Derivation of discrete systems}

We introduce the discrete Lax pair
\begin{equation} \label{Dih-ddLP}
\Psi_{10} = M(p, q_{10}, f;u) \Psi, \quad \Psi_{01} = M(p, q_{01}, g;v) \Psi, 
\end{equation}
where matrix $M$ is given in (\ref{Dih-M}). The compatibility condition of this
Lax pair leads to a set of equations for $p$,  $q$,  $f$ and $g$, 
\begin{subequations} \label{Dih-Dis-CC}
\begin{eqnarray}
&& f_{01} - f - g_{10}+ g + p_{01}\, q_{01}\, -\, p_{10}\, q_{10} =0\, ,
\label{Dih-Dis-CC-1}\\ 
&& (f_{01}-g_{10}) p + g \, p_{01}- f\,  p_{10} + q_{01}-q_{10}= 0,
\label{Dih-Dis-CC-2}\\
&& (f-g) q_{11} + g_{10}\, q_{10} - f_{01}\, q_{01} - p_{01}+ p_{10}=0\, , 
\label{Dih-Dis-CC-3}\\
&& f_{01}\, g - f\, g_{10} + p\, (p_{01}-p_{10}) +q_{11}\,  (q_{01}-q_{10})=0,
\label{Dih-Dis-CC-4}
\end{eqnarray}
\end{subequations}
and an equation solely for $u$ and $v$, 
\begin{equation} \label{Dih-Dis-CC-uv}
 u_{01} v - v_{10}u = 0.
\end{equation}
Functions $u$,  $v$ are apparently redundant since they are completely separated
from the remaining ones and are involved only in equation (\ref{Dih-Dis-CC-uv}).
Taking the value of the first integral $\Phi^1$ in (\ref{dih-D-fi}) to be 1, 
then we can set
$$u^2 \, =\, \frac{1}{f-pq_{10}}, \quad v^2\, =\, \frac{1}{g-p q_{01}}.$$
In view of this substitution,  equation (\ref{Dih-Dis-CC-uv}) becomes
$$ ({\cal{T}}-1)\ln\left(f- p\,  q_{10}\right)\, =\, ({\cal{S}}-1)\ln\left(g-p\,
 q_{01}\right), $$
which can be easily verified to be a conservation law for equations
(\ref{Dih-Dis-CC}). 

Equations (\ref{Dih-Dis-CC}) can be easily solved with respect to $(p_{01},
q_{01}, f_{01}, g)$ or $(p_{10}, q_{10}, f, g_{10})$ leading to a solution with
two branches: the trivial branch (\ref{triv-sol}) and the non-trivial one which
we consider as a system of difference equations. For the latter system it can be
easily verified that it admits two first integrals
\begin{equation} \label{Dih-DD-fi}
({\cal{T}}-1) \frac{f\, -\, p\, q_{10}}{f^2- \left(p^2 + q_{10}^{\, 2}\right)
+1} =0, \quad ({\cal{S}}-1) \frac{g\, -\, p\, q_{01}}{g^2- \left(p^2 +
q_{01}^{\, 2}\right) + 1}=0, 
\end{equation}
a conservation law
$$({\cal{T}}-1)\left(f+p\,  q\right)\, =\, ({\cal{S}}-1)\left(g+p\,  q\right),
$$
and a symmetry given by
\begin{eqnarray*}
\partial_x p &=& (p_{10} q_{10}-p\, q) p + (p-p_{10}) f + q- q_{10},  \\
\partial_x q &=& (p\,  q-p_{-10}q_{-10}) q  + p_{-10}- p + (q_{-10}-q) f_{-10},
\\
\partial_x f &=& (p_{10} q_{10}-p\, q) f + (p-p_{10})p +  (q-q_{10}) q_{10}, \\
\partial_x g &=& (p_{01} q_{01}-p\, q) g + (p-p_{01})p +  (q-q_{01}) q_{01}.
\end{eqnarray*}

Now,  we will consider two particular values for the first integrals
(\ref{Dih-DD-fi}) which allow us to reduce the number of functions involved in
system (\ref{Dih-Dis-CC}) by expressing $f$,  $g$ polynomially in terms of $p$
and $q$. 

\subsubsection{First reduction and a Toda type equation}

Let us first consider for the first integrals the values
$$\frac{f\, -\, p\, q_{10}}{f^2- \left(p^2 + q_{10}^{\, 2}\right) +1} =0, \quad 
\frac{g\, -\, p\, q_{01}}{g^2- \left(p^2 + q_{01}^{\, 2}\right) + 1} =
\frac{1}{2}, $$
which imply that
\begin{equation} \label{Dih-val-fi-0}
f\, -\, p\, q_{10}\, =\, 0, \quad (g-p+q_{01}-1)(g+p-q_{01}-1)\, =\, 0.
\end{equation}
From these algebraic equations,  we choose the solution 
\begin{equation} \label{Dih-val-fi-1}
f\, =\, p\, q_{10}, \quad g\, =\, p-q_{01}+1.
\end{equation}
If we substitute these expressions into system (\ref{Dih-Dis-CC}),  its
conservation laws and symmetry and then make the point transformation $(p, q) =
(\tilde{p}-1, \tilde{q}-1)$,  we will come up with the system
\begin{equation} \label{Dih-red-1-sys}
\tilde{p}_{01}=\frac{\tilde{p}_{10} \tilde{q}_{10}}{\tilde{q}_{11}}, \quad
\tilde{q}_{01}=\frac{(\tilde{p}-2) (\tilde{q}_{10}-2)
\tilde{q}_{11}}{\tilde{p}_{10} \tilde{q}_{10}-2 \tilde{q}_{11}}\, +\, 2, 
\end{equation}
along with its conservation laws
$$({\cal{T}}-1) (\tilde{p}-1) (\tilde{q}_{10}+\tilde{q}-2) = ({\cal{S}}-1) (
\tilde{q} (\tilde{p}-1) - \tilde{q}_{01}), \quad ({\cal{T}}-1) \ln \tilde{p}
\tilde{q}_{10} = ({\cal{S}}-1) \ln \tilde{p} $$
and its symmetry
$$\partial_x \tilde{p} = \tilde{p} (\tilde{p}-2) (\tilde{q}_{10}-\tilde{q}),
\quad \partial_x \tilde{q} = \tilde{q} (\tilde{q}-2)
(\tilde{p}-\tilde{p}_{-10}).$$
$\phantom{m}$

\noindent {\bf{Remark.}} Using the second conservation law above to introduce a potential $w$ by
$$ \tilde{p} \, =\, \exp\left(w-w_{0, -1}\right), \quad \tilde{q}\, =\,
\exp\left(w_{0, -1}-w_{-10}\right), $$
we derive the scalar equation
\begin{equation} \label{Dih-Toda}
{\rm{e}}^{w_{01}-w} - {\rm{e}}^{w-w_{0, -1}}- {\rm{e}}^{w_{1, -1}-w}
+{\rm{e}}^{w-w_{-11}}   =
\frac{1}{2}\Big({\rm{e}}^{w_{01}-w_{-11}}-{\rm{e}}^{w_{1, -1}-w_{0-1}}\Big)
\end{equation}
and its symmetry
$$\partial_x w\, = \, {\rm{e}}^{w-w_{0, -1}}\, -\, {\rm{e}}^{w_{1, -1}-w} \, -\,
\frac{1}{2}\,  {\rm{e}}^{w_{1, -1}-w_{0, -1}}.$$

\subsubsection{Second reduction and a seven point scalar equation}

Another choice for the values of the first integrals (\ref{Dih-DD-fi}) is
$$\frac{f\, -\, p\, q_{10}}{f^2- \left(p^2 + q_{10}^{\, 2}\right) +1} =
\frac{-1}{2}, \quad  \frac{g\, -\, p\, q_{01}}{g^2- \left(p^2 + q_{01}^{\,
2}\right) + 1} = \frac{1}{2}, $$
or,  equivalently, 
\begin{equation} \label{Dih-val-fi-2-0}
(f+p+q_{10}+1) (f-p-q_{10}+1)\, =\, 0, \quad (g-p+q_{01}-1)(g+p-q_{01}-1)\, =\,
0.
\end{equation}
The above equation has four solutions of solutions and we choose 
\begin{equation} \label{Dih-val-fi-2}
f\, =\, p+q_{10}-1, \quad g\, =\, p-q_{01}+1.
\end{equation}
As before,  the substitution of (\ref{Dih-val-fi-2}) into equations
(\ref{Dih-Dis-CC}) and the point transformation $(p, q) \mapsto (\hat{p}+1,
\hat{q}+1)$ result to
\begin{equation} \label{Dih-red-2-sys}
\hat{p}_{01} = \hat{p}_{10}-\hat{q}_{11}+2 + \frac{\hat{p}_{10}
(\hat{q}_{10}-2)}{\hat{p}}, \quad \hat{q}_{01} = \hat{p}\, \frac{\hat{p}_{10}
\hat{q}_{10}-2 \hat{q}_{11}}{\hat{p}_{10} (\hat{p}+\hat{q}_{10}-2)-\hat{p}
\hat{q}_{11}}. 
\end{equation}
Similarly,  we find two conservation laws
\begin{equation} \label{Dih-red-2-sys-cl}
({\cal{T}}-1) \hat{p} \hat{q} = ({\cal{S}}-1) (\hat{p} \hat{q} - 2
\hat{q}_{01}), \quad ({\cal{T}}-1) \ln (\hat{p}-2) (\hat{q}_{10}-2) =
({\cal{S}}-1) \ln \hat{p} (\hat{q}_{01}-2), 
\end{equation}
and a symmetry\footnote{These differential-difference equations are related to
the relativistic Volterra lattice \cite{SR}
$$\partial_y P \, =\,  P \left(Q_{10}-Q + h P_{10} Q_{10} - h P Q\right), \quad
\partial_y Q \, =\, Q \left(P-P_{-10} + h P Q - h P_{-10} Q_{-10} \right)$$
by the point transformation $\left(x, \hat{p}_{ij}, \hat{q}_{ij}\right)\,
\mapsto\, \left(y, P_{ij}, Q_{ij} \right) \, :=\, \left(4\, h\, x,
\frac{\hat{p}_{ij}-2}{2 h}, \frac{\hat{q}_{ij}-2}{2 h} \right)$.}
\begin{equation} \label{Dih-red-2-sys-sym}
\partial_x \hat{p} = (\hat{p}-2) (\hat{p}_{10} (\hat{q}_{10}-2)- \hat{p}
(\hat{q}-2)), \quad \partial_x \hat{q} = (\hat{q}-2) (\hat{q} (\hat{p}-2) -
\hat{q}_{-10} (\hat{p}_{-10}-2)). 
\end{equation}
$\phantom{n}$

\noindent {\bf{Remark.}} Using the first conservation law in (\ref{Dih-red-2-sys-cl}),  we can introduce
a potential $y$ by
\begin{equation} \label{Dih-red-2-y-pot}
\hat{p} \, =\, 2\, \frac{y- y_{-10}}{y_{0, -1}-y_{-10}}, \quad \hat{q}\, =\, 2
(y_{0, -1}-y_{-10}),  
\end{equation}
and derive the Toda-type equation\footnote{If we reverse the $m$ direction, 
then this equation will become equation (D) in \cite{A-JNMP},  cf. also
\cite{A-JPA}.}
\begin{equation}\label{Dih-y-eq}
\left({\cal{S}}-1\right) \log\left(y-y_{-10}\right)\, -\,
\left({\cal{T}}-1\right) \log\left(y_{0, -1}-y\right)\, +\,
\left({\cal{S}}{\cal{T}}^{-1}-1\right) \log\left(1 -
\frac{1}{y-y_{-11}}\right)\, =\, 0
\end{equation}
A conserved form of equation (\ref{Dih-y-eq}) is given by
$$({\cal{S}}-1) \ln \frac{\left(y_{0, -1}-y\right)\left(y_{1,
-1}-y-1\right)}{y_{-10}-y_{0, -1}} = ({\cal{T}}-1) \ln
\frac{\left(y_{-10}-y\right)\left(y_{-11}-y+1\right)}{y_{-10}-y_{0, -1}}, $$
while the differential-difference equation
$$\partial_x y\, =\, (y_{10}-y) (y_{0, -1}-y) \left(\frac{1}{y_{1,
-1}-y}-1\right)\, \equiv\, -(y_{01}-y)
(y_{-10}-y)\left(\frac{1}{y_{-11}-y}+1\right)$$
defines a symmetry of this equation. \\

\noindent {\bf{Remark.}} If we use the second conservation law in (\ref{Dih-red-2-sys-cl}) to introduce
a potential $\phi$ by the relations
\begin{equation} \label{Dih-red-2-f-pot}
\hat{p}\, =\, \frac{2}{1 + {\rm{e}}^{\phi_{0, -1}-\phi_{-10}}}\, \, , \quad
\hat{q}\, =\, 2\, \left( 1 \, +\, {\rm{e}}^{\phi_{0, -1}-\phi_{-1-1}}\, +\,
{\rm{e}}^{\phi_{-10} - \phi_{-1-1}} \right)\,,
\end{equation}
then  system (\ref{Dih-red-2-sys}) will reduce to
equation\footnote{If we reverse the $m$ direction,  the resulting equation will
be equation (E) in \cite{A-JNMP}.}
\begin{equation} \label{Dih-f-eq}
\left({\cal{S}} -1 \right) {\rm{e}}^{\phi-\phi_{-10}}\, +\,
\left({\cal{T}}-1\right) {\rm{e}}^{\phi-\phi_{0, -1}}\, -\, \left({\cal{S}}
{\cal{T}}^{-1} - 1 \right) \frac{1}{1 + {\rm{e}}^{\phi-\phi_{-11}}}\, =\, 0.
\end{equation}
A symmetry of this equation is generated by
$$\partial_x \phi\, =\, {\rm{e}}^{\phi_{01}-\phi}\, -\,
{\rm{e}}^{\phi-\phi_{-10}}\, +\, \frac{1}{1+ {\rm{e}}^{\phi-\phi_{-11}}}\,
\equiv\,  1\, +\, {\rm{e}}^{\phi-\phi_{0, -1}}\, -\, {\rm{e}}^{\phi_{10}-\phi}\,
-\, \frac{1}{1+ {\rm{e}}^{\phi-\phi_{1, -1}}} \, . $$


\noindent {\bf{Remark.}} Combining transformations (\ref{Dih-red-2-y-pot}) and (\ref{Dih-red-2-f-pot}),
 we derive the duality transformation \cite{A-JNMP}
\begin{equation} \label{Dih-y-f-BT}
y_{10}\, -\, y\, =\, {\rm{e}}^{\phi-\phi_{0, -1}}\, +\,  \frac{1}{1 +
{\rm{e}}^{\phi_{1, -1}-\phi}}\, \, , \quad y_{01}\, -\, y\, =\, -\, 
{\rm{e}}^{\phi - \phi_{-10}}\, -\, \frac{1}{1 + {\rm{e}}^{\phi_{-1, 1}-\phi}}\,
\, ,  
\end{equation}
which connect solutions of equations (\ref{Dih-y-eq}),  (\ref{Dih-f-eq}).

\section{Concluding remarks}

In this paper we discussed the Darboux-Lax scheme for Lax operators related to nonlinear NLS type equations. 
We derived integrable systems of differential-difference and partial difference equations, and discussed several reductions to 
scalar Toda-type equations. The results of this paper have already been employed in other works. For some of the systems presented here the symmetry 
structure and recursion operators were studied in   \cite{KMW}, whereas  the connection of 
our systems with Yang-Baxter maps was explored in \cite{KM}. Lax-Darboux schemes corresponding to 
other Lie algebras are discussed in \cite{BMX, MPW}.

\section*{Acknowledgements}
This work initiated when AM and PX were participating in the programme
``{\em{Discrete Integrable Systems}}'' at the Isaac Newton Institute, 
Cambridge,  UK in 2009. PX was initially supported by the Newton International Fellowship grant NF082473. 
AM and PX gratefully acknowledge support from an EPSRC grant EP/I038675/1. 
AM gratefully acknowledges support from a Leverhulm Trust grant.
SKR would like to thank University of Leeds for the William Right Smith
Scholarship,  and John E. Crowther for the Scholarship-Contribution to Fees.

\end{document}